\documentclass[onecolumn,preprintnumbers,amsmath,amssymb,
groupedaddress,prc,floatfix]{revtex4}
\usepackage{mathrsfs}
\usepackage{longtable}
\usepackage{graphicx}
\usepackage{epsfig}
\usepackage{dcolumn}
\usepackage{subfigure}
\usepackage{rotating}
\begin{document}


\title{Spectroscopy of $^{196}$Hg using Digital INGA at VECC, Kolkata}

\author{S. Das}
\author{S. Chatterjee}
\author{S. Samanta}
\author{R. Raut}
\author{S. S. Ghugre}
\address{UGC-DAE Consortium for Scientific Research, Kolkata Centre, Kolkata 700098, India}

\author{A. K. Sinha}
\address{UGC-DAE Consortium for Scientific Research, Indore 452001, India}

\author{S. Bhattacharya}
\author{S. Nandi}
\author{R. Banik}
\author{G. Mukherjee}
\author{S. Bhattacharyya}
\address{Physics Group, Variable Energy Cyclotron Centre, Kolkata 700064, India}
\address{Homi Bhaba National Institute, Training School Complex, Anushaktinagar, Mumbai-400094, India}

\author{S. Ali}
\author{P. Ray}
\author{H. Pai}
\author{A. Goswami}
\address{Nuclear Physics Division, Saha Institute of Nuclear Physics, Kolkata 700064, India}

\author{R. Rajbanshi}
\address{Presidency University, 86, 1, College St, Kolkata, West Bengal 700073}

\author{S. Das Gupta}
\address{Victoria Institution (College), Kolkata 70000, India}


\maketitle

\section{Introduction}

\begin{figure}
\includegraphics[angle=0,scale=0.45,trim=5.0cm 1.0cm 6.0cm 1.0cm,clip=true]{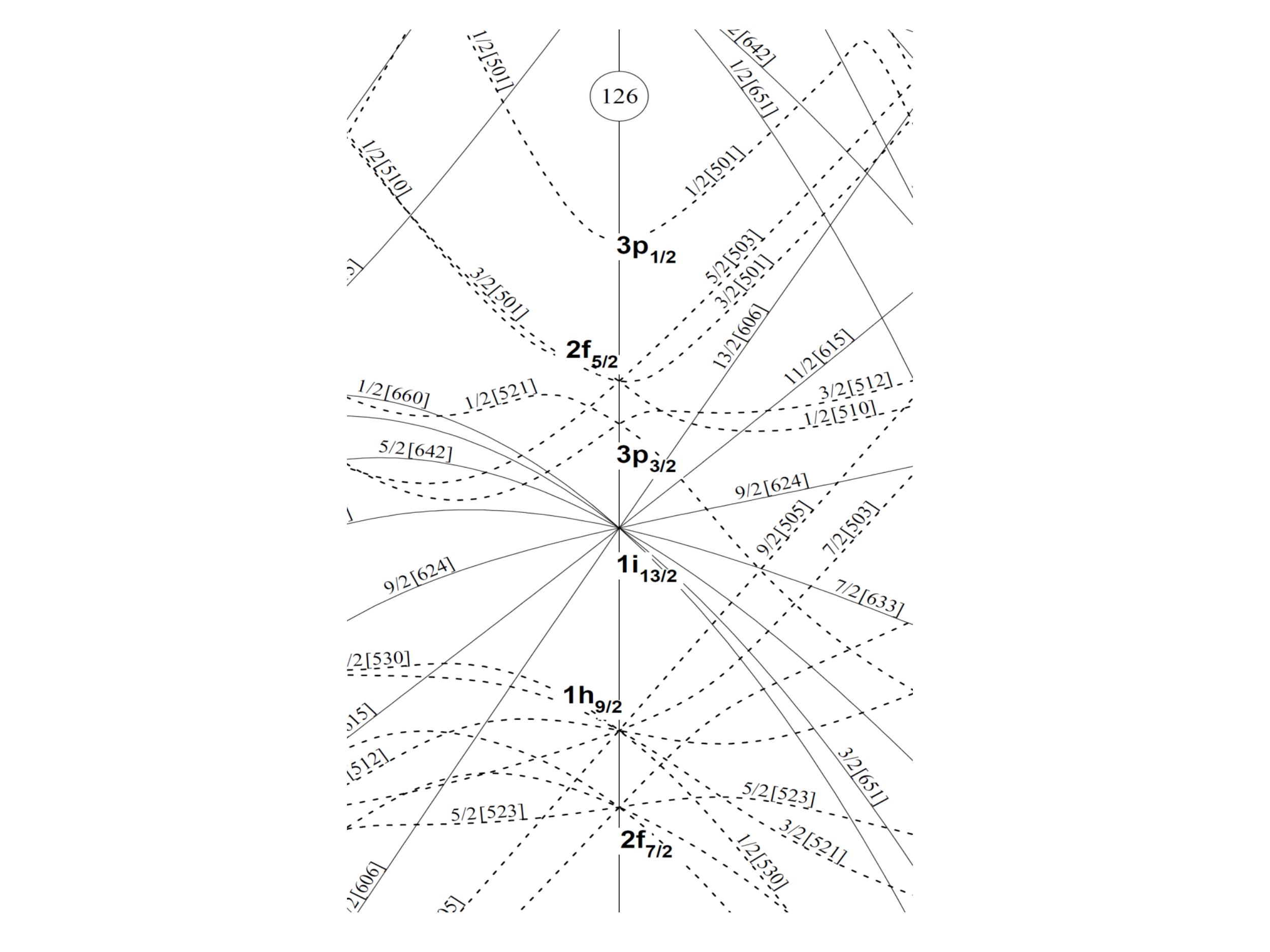}
\includegraphics[angle=0,scale=0.45,trim=6.0cm 1.0cm 5.0cm 1.0cm,clip=true]{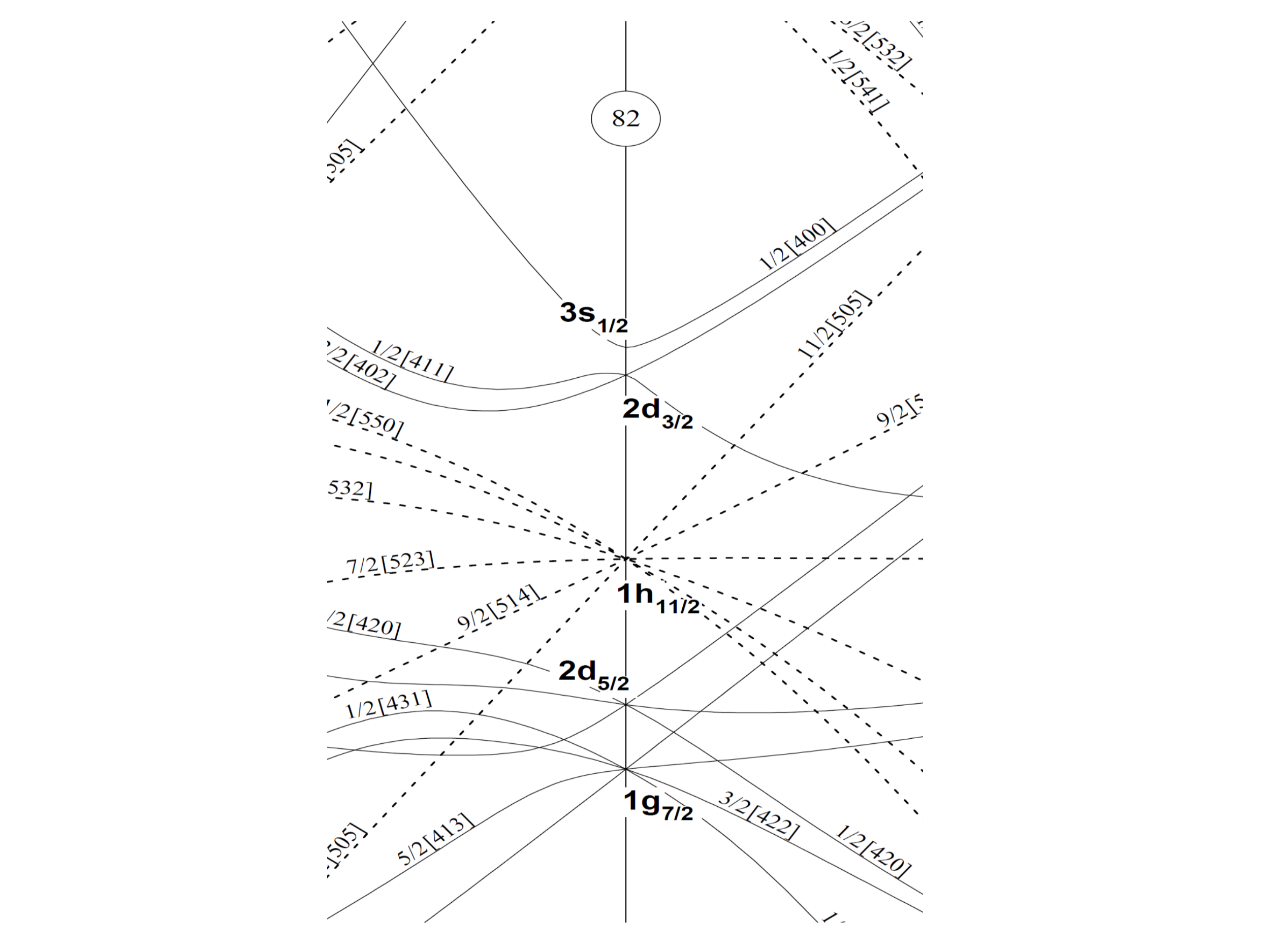}
\caption{\label{fig:4.1} Relevant orbitals for neutrons (left panel) and protons (right panel) underlying the excitation scheme of $^{196}$Hg nucleus.}
\end{figure}

The low and high spin states in Hg isotopes with A $\sim$ 190-200 have been extensively studies over the last few decades, as discussed subsequently. These
isotopes, with two protons less than the Z=82 magic number, lie in the transitional region between the strongly deformed prolate rare-earth and the spherical
Pb nuclei. There are several features common to the Hg isotopes in the mass range 190-200. The ground state bands of the even-A isotopes have a $2^{+}$ state
at excitation $\sim$ 420 keV. The E2 de-excitation therefrom is an enhanced one. The $2^{+}$ state has been interpreted, in the framework of particle-vibrator
coupling scheme, as a two proton-hole \textquotedblleft{intruder}\textquotedblright{} state with configuration (${\pi}s_{1/2}^{-1}{\pi}d_{3/2}^{-1}$) \cite{Lie75}
(Fig. \ref{fig:4.1}). The ground state band in the even Hg isotopes further consists of $4^{+}, 6^{+}, 8^{+}$ sequence that are approximately equidistant
\cite{Gun77} followed by an isomeric $10^{+}$ level with half-life $\sim$ 7-25 ns \cite{Lie75}. The $4^{+}$ and $6^{+}$ states have been interpreted as phonon
multiplet states while the $8^{+}$ and the $10^{+}$ as belonging to two proton-hole (${\pi}h_{11/2}^{-2}$) configuration. Compression in the level energies and
brunching of states have been observed around the $10^{+}$ level and the same could be explained by particle-vibration coupling model \cite{Gun77}. The even-A
Hg isotopes also exhibit a negative parity band, consisting of states $5^{-}, 7^{-}, 9^{-},....$ \cite{Lie75}, that de-excite into the ground state band. This
has been interpreted to be of collective origin based on a neutron in i$_{13/2}$ and one in the adjacent p$_{1/2}$, p$_{3/2}$ or f$_{5/2}$ orbital. The level
structures of odd-A Hg isotopes, particularly in the A $\sim$ 190-200 range, share an intriguing overlap with the neighboring even-A ones. The ground state
sequence in the odd Hg isotopes is constituted with $13/2^{+}, 17/2^{+}, 21/2^{+}....$ states wherein the transition energies are very similar to those in the
ground state band of the adjacent even-A isotopes. This sequence in odd-A Hg isotopes has been interpreted as based on a neutron in $i_{13/2}$ orbital decoupled
from a weakly deformed oblate even-A (Hg) core \cite{Lie75}. Further, negative parity sequences, de-exciting to the ground state bands, have also been observed
in the odd Hg isotopes. The intraband transition energies therein overlap with those in the analogous negative sequences, based on the $5^{-}$ state, of the
adjacent even-A Hg nuclei. These negative parity bands in odd Hg isotopes have been interpreted, as based on three quasi neutrons decoupled from the core
\cite{Lie75}. The aforementioned findings on the low-spin level structures of Hg isotopes were typically based on experimental investigations carried out
using $\alpha$-induced reactions and modest detection setups consisting of few Ge detectors \cite{Pro74,Lie75,Gun77}. Eventually, the structural studies in Hg
isotopes have been extended to higher spin domain wherein updated and developed experimental facilities were used \cite{Ced93}.\\

The spectroscopic investigation of $^{196}$Hg (Z=80, N=116), carried out as a part of the current thesis, was primarily addressed to the low spin excitation
pattern of the nucleus. The nucleus was previously studied by a number of researchers such as Proetel \textit{et al.} \cite{Pro74} who carried out $\gamma$-ray
spectroscopy of $^{196}$Hg using $\alpha$-induced reaction to populate the same with a couple Ge(Li) detectors as the detection system. The ground state (positive
parity) band was identified upto an excitation energy (E$_{x}$) $\sim$ 2.7 MeV and spin-parity ($J^{\pi}$) $\sim 12^{+}$ while the negative parity sequence was
confirmed upto E$_{x}$ $\sim$ 2.6 MeV and $J^{\pi}$ $\sim 11^{-}$. These bands were considerably extended by Helppi \textit{et al.} \cite{Hel83} following the
population of $^{196}$Hg in $^{196}$Pt($\alpha$,4n) reaction at E$_{lab}$ $\sim$ 49 MeV and using 1-2 Ge(Li) detectors for detecting the emitted $\gamma$-rays.
The ground state band was extended upto E$_{x}$ $\sim$ 4.3 MeV and $J^{\pi}$ $\sim 18^{+}$. The negative parity sequence was also extended to E$_{x}$ $\sim$
5 MeV and $J^{\pi}$ $\sim 21^{-}$. The aligned angular momenta therein were found to be in overlap with the theoretical estimates. The nucleus was latter
investigated using improved experimental setup by Mehta \textit{et al.} \cite{Meh91}. The spectroscopy therein was carried out following the production of the
($^{196}$Hg) nucleus in $^{198}$Pt($\alpha$,6n) reaction at E$_{lab}$ $\sim$ 78 MeV (degraded from 90 MeV) and using an array of six Compton suppressed Ge
detectors as the detection system. Several bands upto E$_{x}$ $\sim$ 6-7 MeV and spin $\sim$ 23-25$\hbar$ were reported therefrom. The experimental observations
were interpreted in the framework of Cranked Shell Model (CSM) calculations and the theoretical results were found in satisfactory compliance with the measurements
when $i_{13/2}$ neutron alignments were assumed. The neutron $i_{13/2}$ configurations underlying the high spin states of $^{196}$Hg was earlier identified by
Kroth \textit{et al.} \cite{Kro81}. High spin states of $^{196}$Hg were later studied by Cederwall \textit{et al.} \cite{Ced93} using $^{192}$Os($^{9}$Be,5n)
reaction at E$_{lab}$ $\sim$ 65 MeV, to populate the nucleus, and an array of 20 Compton suppressed Ge detectors, as the detection system. The latter was
augmented with a 4$\pi$ array of 40 BGO detectors for selection of higher multiplicity events. A high spin band extending upto E$_{x}$ $\sim$ 8.7 MeV and spin
$\sim$ 30$\hbar$ was identified in the process. The structure of the $^{196}$Hg nucleus was investigated later, albeit in a different context, by Bernards
\textit{et al.} \cite{Ber10} using $^{194}$Pt($\alpha$,2n) reaction at E$_{lab}$ = 9.3 MeV and an array of 13 HPGe detectors as the detection system.\\

As far as the present work is concerned, the same was aimed at observing new low spin non-yrast structures in $^{196}$Hg and, in the process, confirm and extend
the existing excitation scheme of the nucleus. The pursuit has been detailed hereafter.

\begin{figure}
\includegraphics[angle=0,scale=0.40,trim=1.0cm 0.5cm 2.0cm 3.0cm,clip=true]{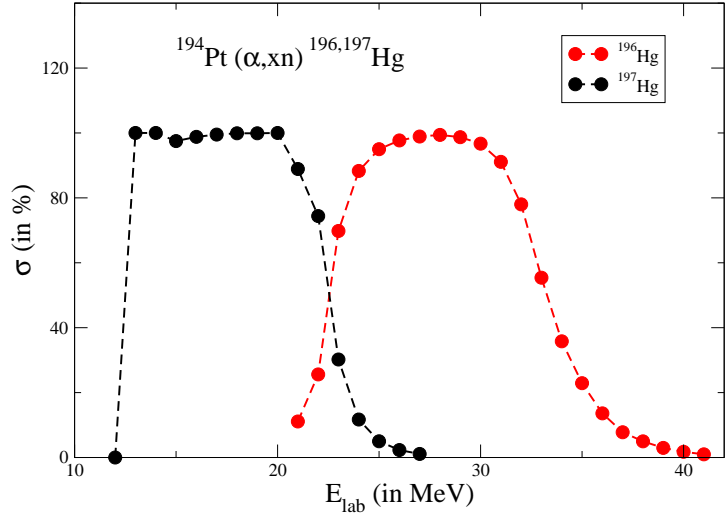}
\caption{\label{fig:4.2}Variation of relative cross-sections ($\sigma$) with the energies of the $\alpha$-beam. Only the most dominant channels are shown here.}
\end{figure}

\section{Details of the Experiment and Data Analysis}

\begin{figure}
\includegraphics[angle=90,scale=0.38,trim=0.0cm 0.0cm 0.0cm 0.0cm,clip=true]{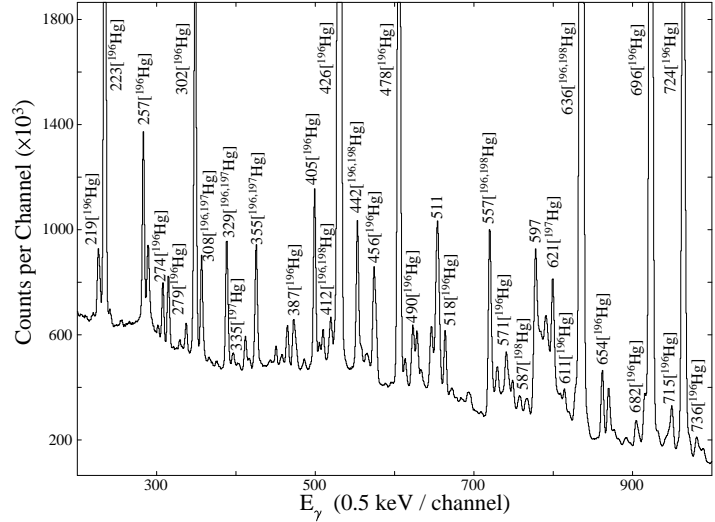}
\caption{\label{fig:4.3}Part of the projection spectrum constructed out of the present data illustrating the copious population of $^{196}$Hg nucleus in the experiment.}
\end{figure}

The nucleus of interest $^{196}$Hg was populated in the present experiment using $^{194}$Pt($\alpha$,2n)$^{196}$Hg reaction at E$_{lab}$ = 32 MeV. The $\alpha$-beam
was obtained from the K-130 room temperature cyclotron facility at the Variable Energy Cyclotron Centre (VECC), Kolkata. The target was 13.6 mg/cm$^{2}$ thick
foil of $^{194}$Pt (enriched to $\sim$ 97\%). The choice of beam energy was based on the statistical model calculations using PACE4 code \cite{Gav80} and was
aimed at maximizing the yield of the nucleus of interest. A plot of the relative cross-sections of $^{196,197}$Hg, for the aforementioned reaction, against
the energies of the $\alpha$-beam, is illustrated in Fig. \ref{fig:4.2}. The $\gamma$-rays from the de-exciting nuclei were detected using the INGA setup at
VECC. During the experiment the array comprised of 7 Compton suppressed Clover detectors placed at 90$^{\circ}$ (4 detectors), 125$^{\circ}$ (2 detectors) and
40$^{\circ}$ (1 detector) angle. The digitizer based pulse processing and data acquisition system was employed which is based on Pixie-16 250-MHz, 12-bit
digitizer modules from XIA LLC, USA \cite{Das18}. In-beam list-mode data were acquired under the trigger condition of atleast two Compton suppressed Clovers firing
in coincidence. Around 5$\times$10$^{8}$ of two and higher fold events were recorded during the experiment. The acquired data was sorted using IUCPIX package
\cite{Das18}.\\

\begin{figure}
\includegraphics[angle=0,scale=0.38,trim=0.5cm 0.5cm 3.0cm 3.0cm,clip=true]{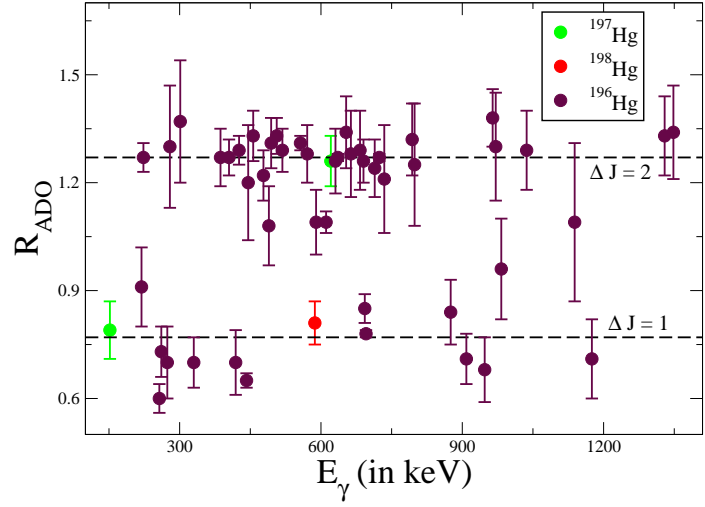}
\caption{\label{fig:4.4}R$_{ADO}$ values for different transitions of $^{196}$Hg along with those of selected transitions of previously known multipolarities from
other nuclei ($^{197,198}$Hg) populated in the present experiment. The latter is used to fix the reference values used in the current analysis as well as for
validation of the same.}
\end{figure}

A part of the raw projection spectrum, constructed from the coincidence data acquired in the current experiment, is illustrated in Fig. \ref{fig:4.3}. The
$\gamma$-ray transition peaks identified in Fig. \ref{fig:4.3} establish the population of $^{196}$Hg following the chosen reaction. As has been mentioned
earlier, the $\alpha$-induced reactions are typically characterized by dominant population of few reaction channels that facilitate them with copious statistics.
It is indicated in Fig. \ref{fig:4.3} that the population of $^{196}$Hg in the present reaction is accompanied by that of $^{197,198}$Hg as well albeit with
much lower dominance. To quantify the same, based on the peak areas of the respective ground state transitions, $^{196}$Hg contribute to $\sim$ 98 \% of the
total yield while the other isotopes make up for the rest. However, owing to the copious population of the limited number of reaction channels in the $\alpha$
-induced reaction, even the 2 \% yield led to sufficient statistics even for R$_{ADO}$ and polarization measurements.\\

\begin{figure}
\includegraphics[angle=0,scale=0.38,trim=0.0cm 0.5cm 3.0cm 3.0cm,clip=true]{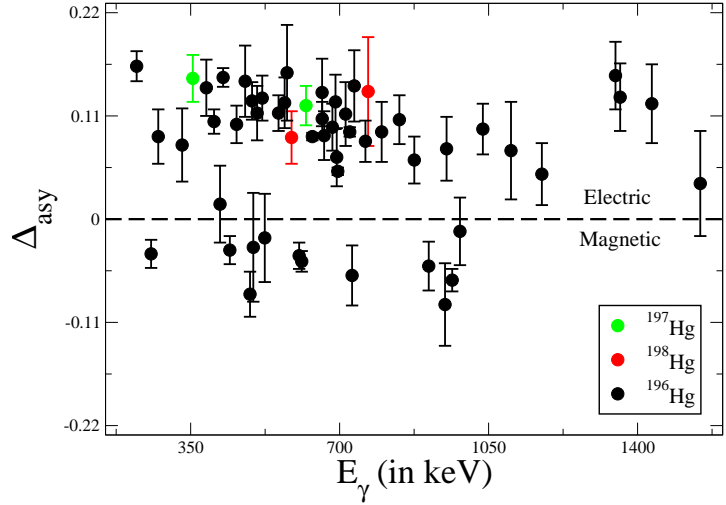}
\caption{\label{fig:4.5}Plot of polarization asymmetry (${\Delta}_{asymm.}$) \cite{Sam18} for different $\gamma$-ray transition in $^{196}$Hg along with those in
other nuclei ($^{197,198}$Hg) produced in the same experiment, included as reference.}
\end{figure}

The acquired data were sorted into symmetric (all vs all) and asymmetric or angle-dependent $\gamma$-$\gamma$ matrices for the subsequent analysis procedures. The
data sorting procedure was carried out, using IUCPIX-package \cite{Das18} and the RADWARE package \cite{rad} was used for the subsequent analysis.\\

The multipolarities of $\gamma$-ray transitions in the present analysis was assigned from their Ratio of Angular Distribution from Oriented nuclei (R$_{ADO}$)
\cite{Sam18} defined here as,
\begin{equation}
R_{ADO} = \frac{I_{\gamma}^{40^{\circ}}}{I_{\gamma}^{90^{\circ}}}
\end{equation}

where, the numerator and the denominator represents intensity of the $\gamma$-ray transition at respective angles, that is in coincidence with correlated
$\gamma$-rays detected in any detector. Two angle dependent matrices were constructed for the purpose. One had $\gamma$-rays detected in any of the detectors at
40$^{\circ}$ on Y-axis while the coincident $\gamma$-rays detected in any other detectors on X-axis. The other had $\gamma$-rays detected in one of the 90$^{\circ}$
detectors on the Y-axis while the coincident $\gamma$-rays in any detector on the X-axis. In the present setup, the expected value of R$_{ADO}$ for pure quadrupole
and pure dipole transitions are 1.27$\pm$.01 and 0.77$\pm$.01, respectively. These were determined from the weighted average of the R$_{ADO}$ values calculated
for $\gamma$-ray transitions of previously known pure multipolarities, that were produced in the experiment. A R$_{ADO}$ value between the values for pure
transitions would indicate transitions of mixed multipolarity (with mixing ratio, $\delta$ $>$ 0). The plot of R$_{ADO}$ values of $\gamma$-ray transitions for the
nucleus of interest, $^{196}$Hg, along with those used to determine the reference values is presented in Fig. \ref{fig:4.4}. The multipolarity assignments based on
these numbers are discussed in the next section.\\

\begin{figure}
\includegraphics[angle=0,scale=0.38,trim=0.5cm 1.0cm 3.0cm 3.0cm,clip=true]{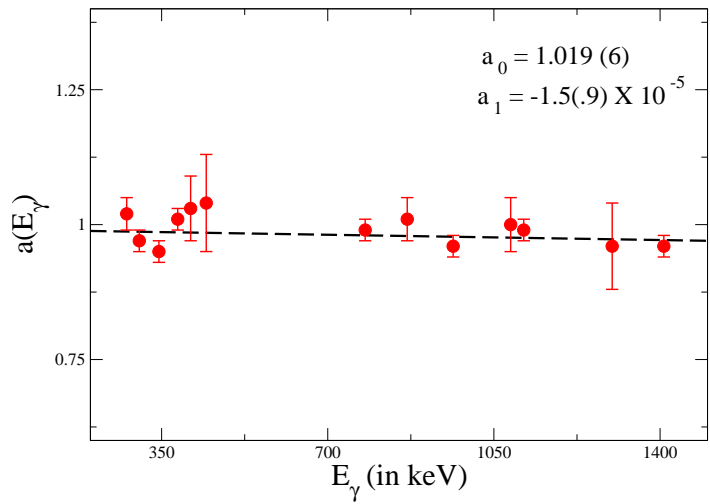}
\caption{\label{fig:4.6}Plot of the geometrical asymmetry ($a$) against $\gamma$-ray energies (E$_{\gamma}$) along with the fit to the data points using $a_{0} 
+ a_{1}$E$_{\gamma}$}
\end{figure}

The use of Clover detectors facilitate extraction of linear polarization information of the observed $\gamma$-rays. The polarization of a $\gamma$-ray transition
is indicative of its electromagnetic (electric or magnetic) character and is determined from the asymmetry (${\Delta}_{asymm.}$) between its scattering in the
perpendicular and the parallel planes with respect to the reaction plane \cite{Sam18}. Two asymmetric matrices were constructed for extraction of the
${\Delta}_{asymm.}$. These had $\gamma$-rays detected by all detectors on the X-axis and those detected in coincidence by the perpendicular (parallel) combination
of crystals in the 90$^{\circ}$ Clover detectors on the Y-axis. The plot of ${\Delta}_{asymm.}$ values of $\gamma$-ray transitions for nucleus of interest $^{196}$Hg
along with those from other nuclei ($^{197,198}$Hg) used to validate the analysis is presented in Fig. \ref{fig:4.5}. A positive value of ${\Delta}_{asymm.}$ is
indicative of an electric nature while a negative value implies that the $\gamma$-ray transition is magnetic. A near-zero usually signifies a mixed electromagnetic
character. The inherent geometrical asymmetry ($a$ \cite{Sam18} ) of the setup was determined using unpolarized $^{152}$Eu and $^{133}$Ba radioactive sources
\cite{Sam18}. The plot of $a$ against $\gamma$-ray energies is illustrated in Fig. \ref{fig:4.6} along with the fit to the data points using
$a_{0} + a_{1}$E$_{\gamma}$. The values of $a_{0}$ = 1.019$\pm$0.006 was used for $a$ in the calculation of ${\Delta}_{asymm.}$, since $a_{1}$ =
(1.5$\pm$0.9)$\times$10$^{-5}$ was perceived to be insignificantly small to impact the ${\Delta}_{asymm.}$ values or the assignments therefrom.\\

\begin{figure}
\includegraphics[angle=0,scale=0.50,trim=0.0cm 0.0cm 1.5cm 1.0cm,clip=true]{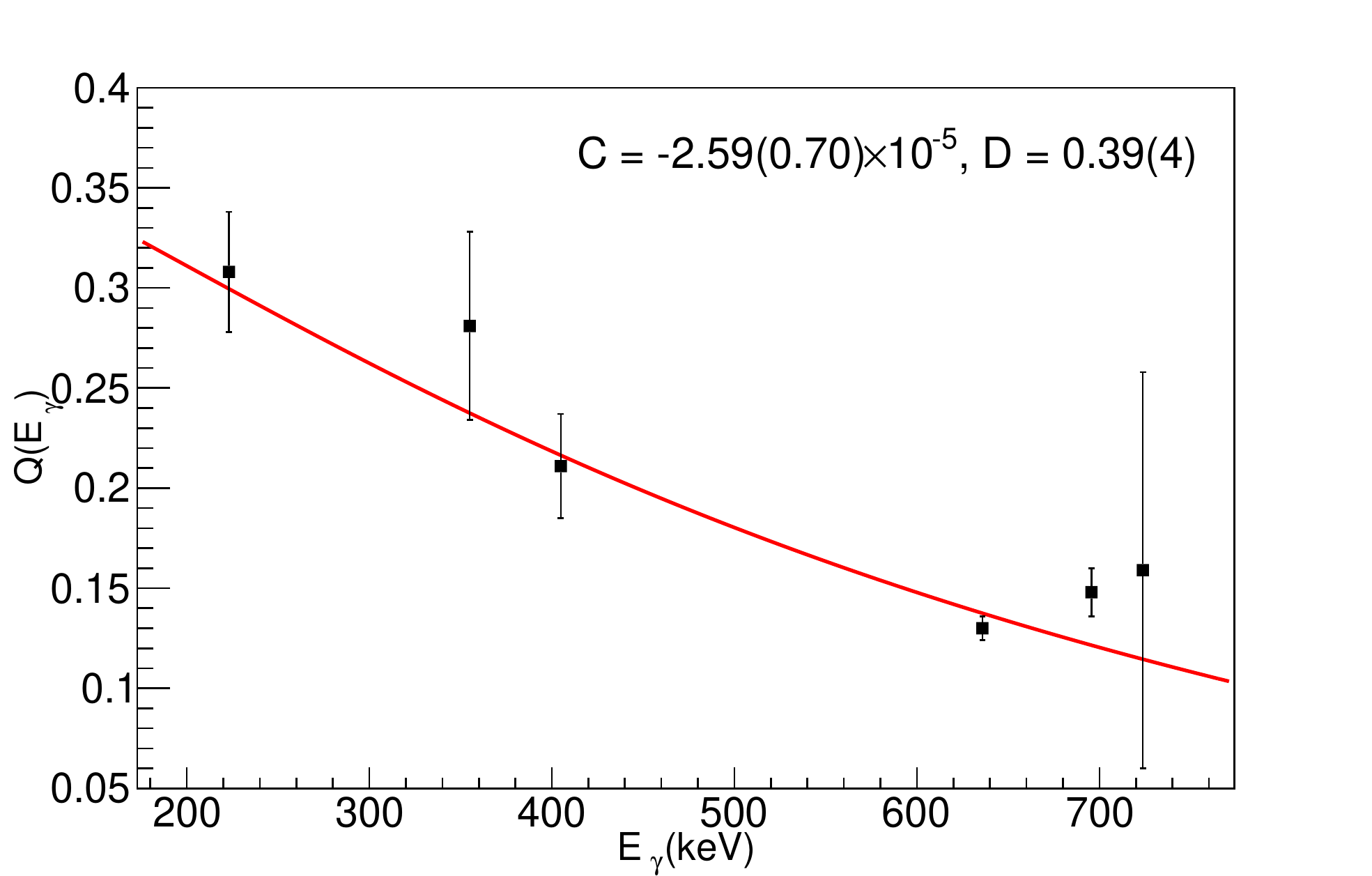}
\caption{\label{fig:4.7}Plot of polarization sensitivity (Q(E$_{\gamma}$)) as a function of $\gamma$-ray energy, determined from the observed $\gamma$-rays of
previously known multipolarities, along with the fitting \cite{Sam18}.}
\end{figure}

\begin{figure}
\includegraphics[angle=0,scale=0.38,trim=0.5cm 1.0cm 2.8cm 3.0cm,clip=true]{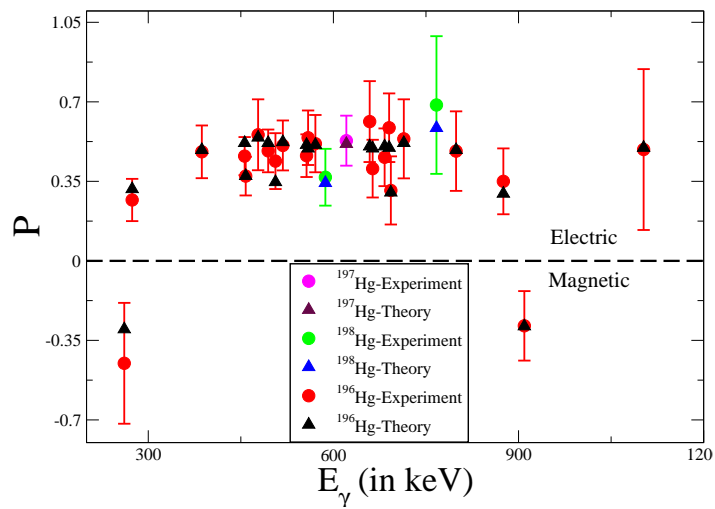}
\caption{\label{fig:4.8}Plot of polarization values, experimental and theoretical, for $\gamma$-rays observed in the present experiment.}
\end{figure}

Further to the calculation of ${\Delta}_{asymm.}$, polarization (P) of the nuclei of interest were also determined \cite{Sam18}. The polarization sensitivity
($Q$) of the setup was calculated from $\gamma$-rays of previously known multipolarities. The plot of $Q$ against $\gamma$-ray energies along with the fit using
Eq. Q = Q$_{0}$(CE$_{\gamma}$ + D) is illustrated in Fig. \ref{fig:4.7}. The parameters C and D obtained therefrom were used to determine the experimental values
of the polarization (P) for the $\gamma$-rays of interest. These are plotted in Fig. \ref{fig:4.8} along with the corresponding theoretical values calculated using
the procedure described in Ref \cite{Pal00}. The analysis was also extended to the previously known transitions of $^{197,198}$Hg as a validation of the present
exercise. The overlap between the experimental and theoretical P values was found to be satisfactory (for electromagnetic assignments).\\

The $\gamma$-rays in $^{196}$Hg identified and confirmed in the present endeavor along with their multipolarity and electromagnetic assignments are recorded
in the next section.

\section{Results and Discussions}

\begin{sidewaysfigure}
\includegraphics[angle=0,scale=1.10,trim=0.0cm 7.0cm 0.0cm 7.0cm,clip=true]{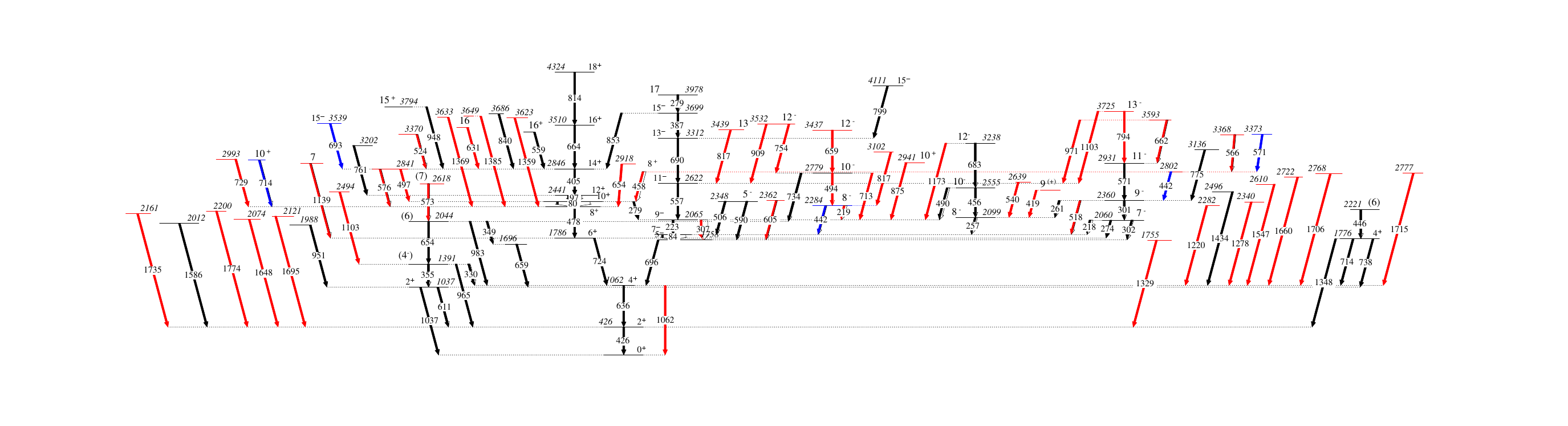}
\caption{\label{fig:4.9}Level scheme of the $^{196}Hg$ nucleus from the present work. The new $\gamma$-ray transitions identified from the current study are
indicated by red whereas the transitions labeled in blue were observed in the previous studies but were either not placed in the level scheme or had different
placement with respect to the energy and/or $J^{\pi}$ values of the de-exciting states.}
\end{sidewaysfigure}

The level scheme of $^{196}$Hg, as established from the current investigation, is illustrated in Fig. \ref{fig:4.9}. The list of levels and $\gamma$-ray
transitions of $^{196}$Hg, along with their spectroscopic properties, are recorded in Table \ref{table:4.1}. The energies of the levels and the $\gamma$-ray
transitions, rounded off to the first decimal, are listed in Table \ref{table:4.1} while the same, approximated to the nearest integer, are labeled in Fig.
\ref{fig:4.9} of the excitation scheme. The subsequent discussion on the level scheme is mostly based on the integer energies included therein unless
otherwise required. The adjacent figures in this section represent typical gated spectra, from $\gamma$-$\gamma$ matrix and $\gamma$-$\gamma$-$\gamma$ cube,
illustrating the coincident $\gamma$-rays constituting the level structure of $^{196}$Hg. The level scheme of the nucleus has been established upto an
excitation energy E$_{x} \sim$ 4.3 MeV and spin $\sim 18 \hbar$. More than 40 new transitions were identified and placed in the scheme in addition to
confirming or modifying the placements of some of the existing ones from previous studies \cite{Meh91,Hel83,Kro81,Gut83}. Since the R$_{ADO}$ measurement
was carried out with a single detector at 40$^{\circ}$, the same was, at times, plagued with sparse statistics on the $\gamma$-ray transition peaks of
interest. Thus, for certain transitions, that are first observed in the present work, while the electromagnetic assignment would be made from measurement
of linear polarization, the multipolarity assignment is missing. This is the maiden instance when electromagnetic assignments have been made directly from
the measurement of linear polarization of the $\gamma$-ray transitions.\\

\begin{figure}
\includegraphics[angle=0,scale=0.30,trim=0.0cm 0.0cm 0.0cm 0.0cm,clip=true]{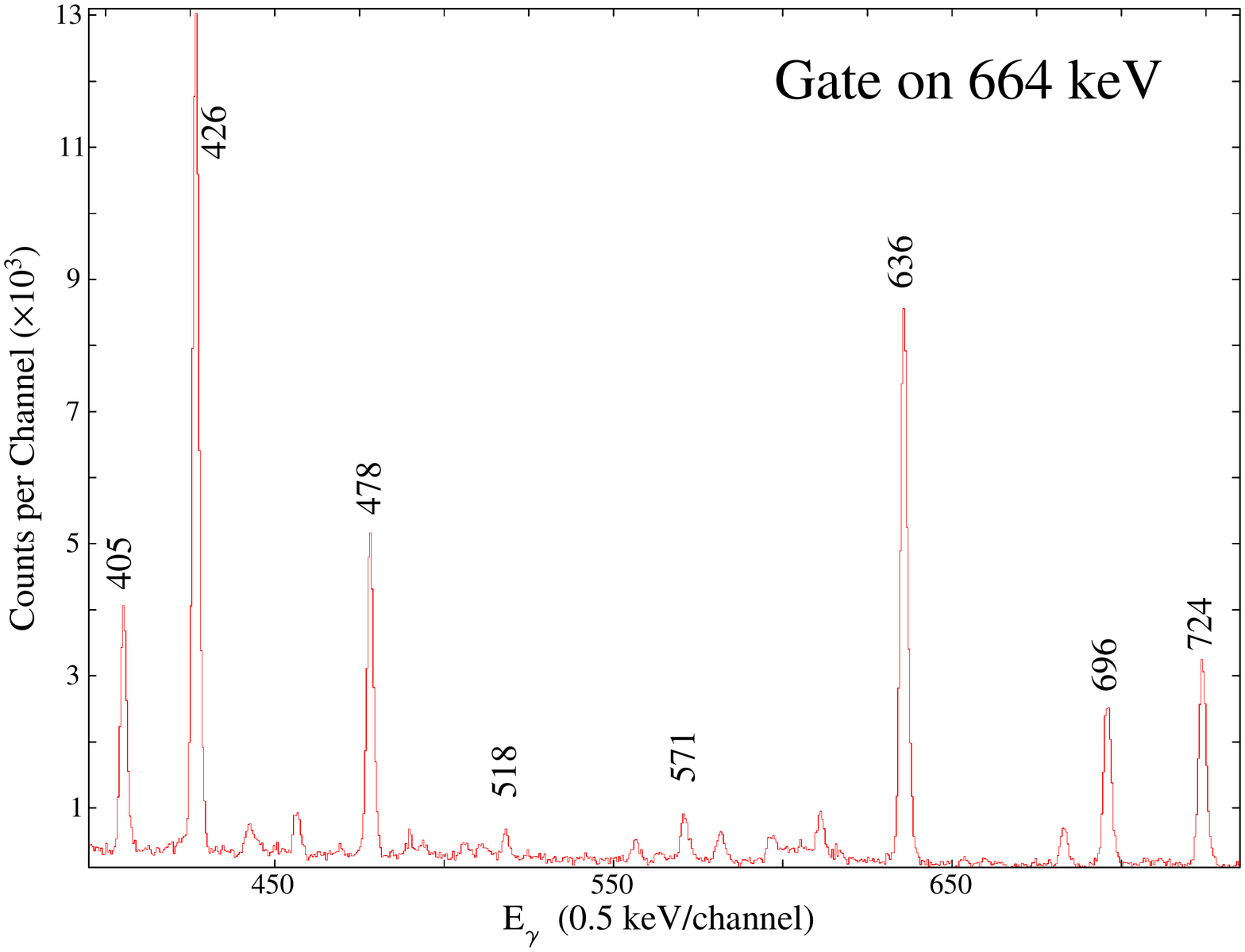}
\includegraphics[angle=0,scale=0.30,trim=0.0cm 0.0cm 0.0cm 0.0cm,clip=true]{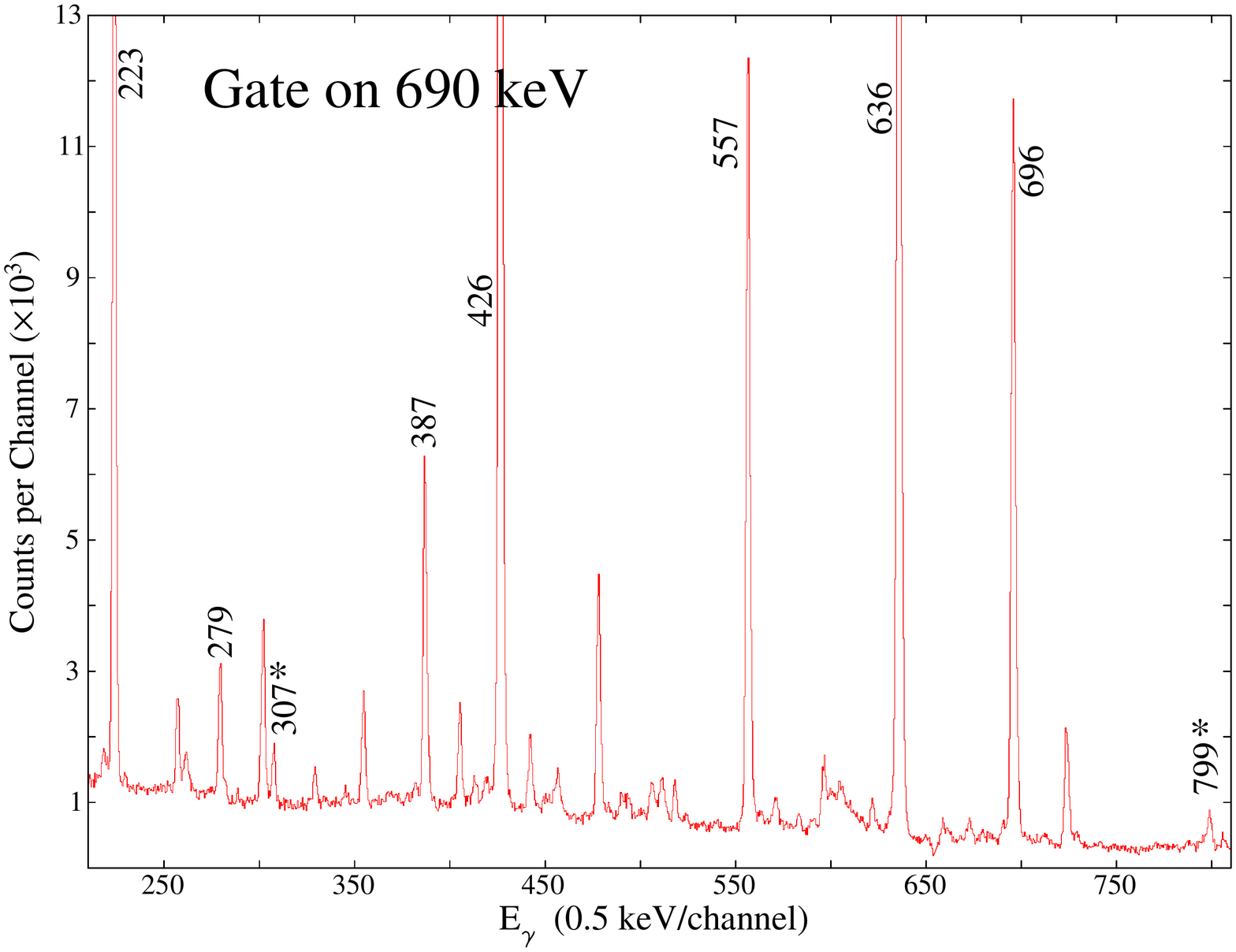}
\caption{\label{fig:4.10}Representative spectra with gate on 664 (for positive parity ground state band depicted in left panel) and 690 keV (for negative parity
yrast band depicted in right panel) transitions of $^{196}Hg$ . The new transitions, first observed in the present study, are labeled with an asterisk (*).}
\end{figure}

\begin{figure}
\includegraphics[angle=90,scale=0.30,trim=0.0cm 0.0cm 0.0cm 0.0cm,clip=true]{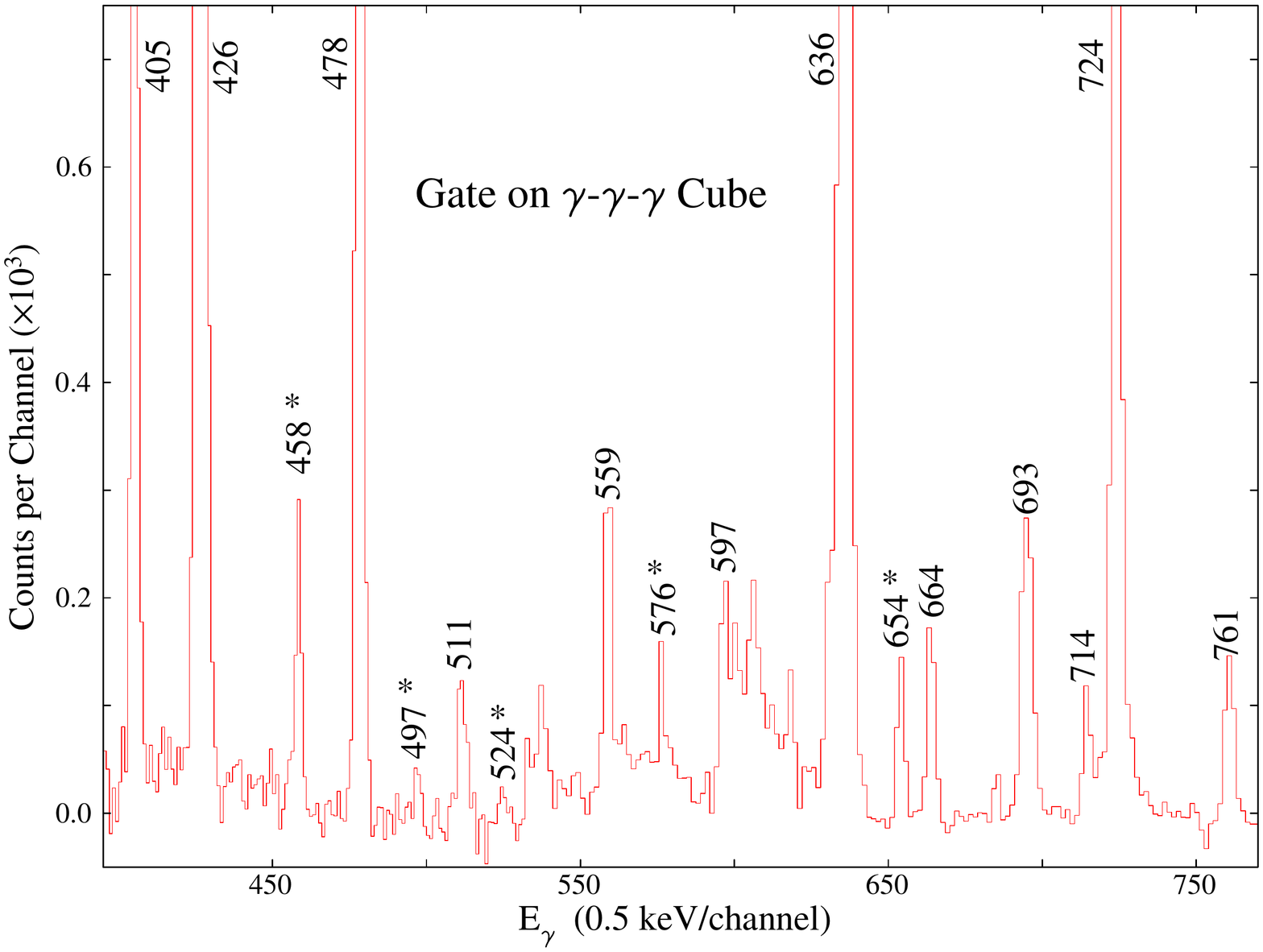}
\includegraphics[angle=90,scale=0.30,trim=0.0cm 0.0cm 0.0cm 0.0cm,clip=true]{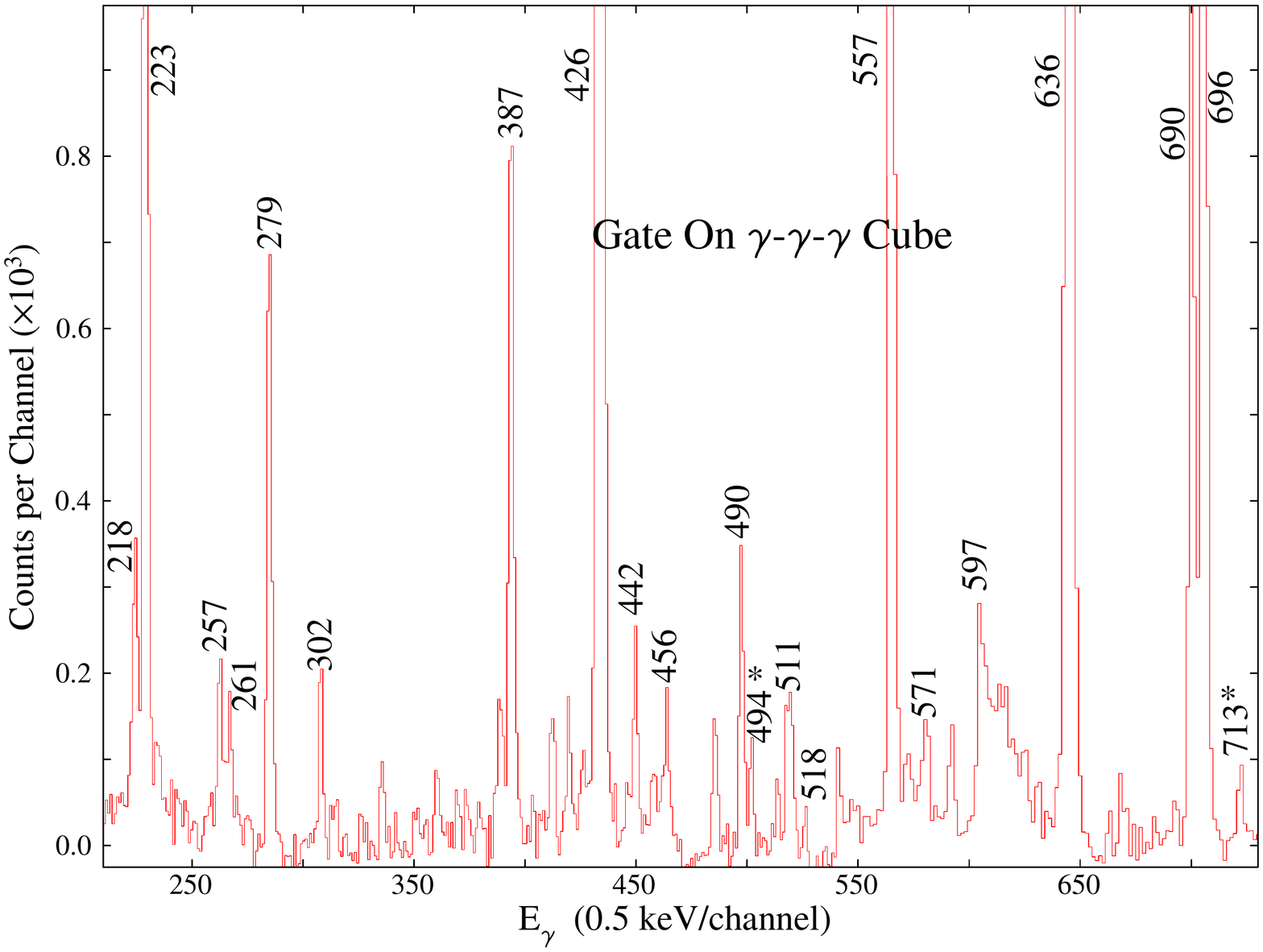}
\caption{\label{fig:4.11}Sum spectra of $\gamma$-$\gamma$ gates, applied on the $\gamma$-$\gamma$-$\gamma$ cube, from a list of transitions, 724, 478, 405,
664, 814 (left panel) and 696, 223, 557, 690, 799 (right panel) that illustrates the $\gamma$-rays of $^{196}Hg$. The new transitions, first observed in the
present study, are labeled with an asterisk (*).}
\end{figure}

The positive parity ground state band and the negative parity yrast band, de-exciting into the ground band, as previously established in $^{196}$Hg could be
partially observed  in the present experiment. The findings therein overlapped with the existing literature on the nucleus and the multipolarities and the
electromagnetic nature of the corresponding intraband $\gamma$-ray transitions were used to validate the current data analysis exercise. Spectra extracted
from both $\gamma$-$\gamma$ matrix and $\gamma$-$\gamma$-$\gamma$ cube, with gate on transitions of these bands are illustrated in Fig. \ref{fig:4.10} and
Fig. \ref{fig:4.11} and uphold the associated placements from previous studies.\\

\begin{figure}
\subfigure[]{\label{fig:4.12a}\includegraphics[angle=0,scale=0.30,trim=0.0cm 0.0cm 0.0cm 0.0cm,clip=true]{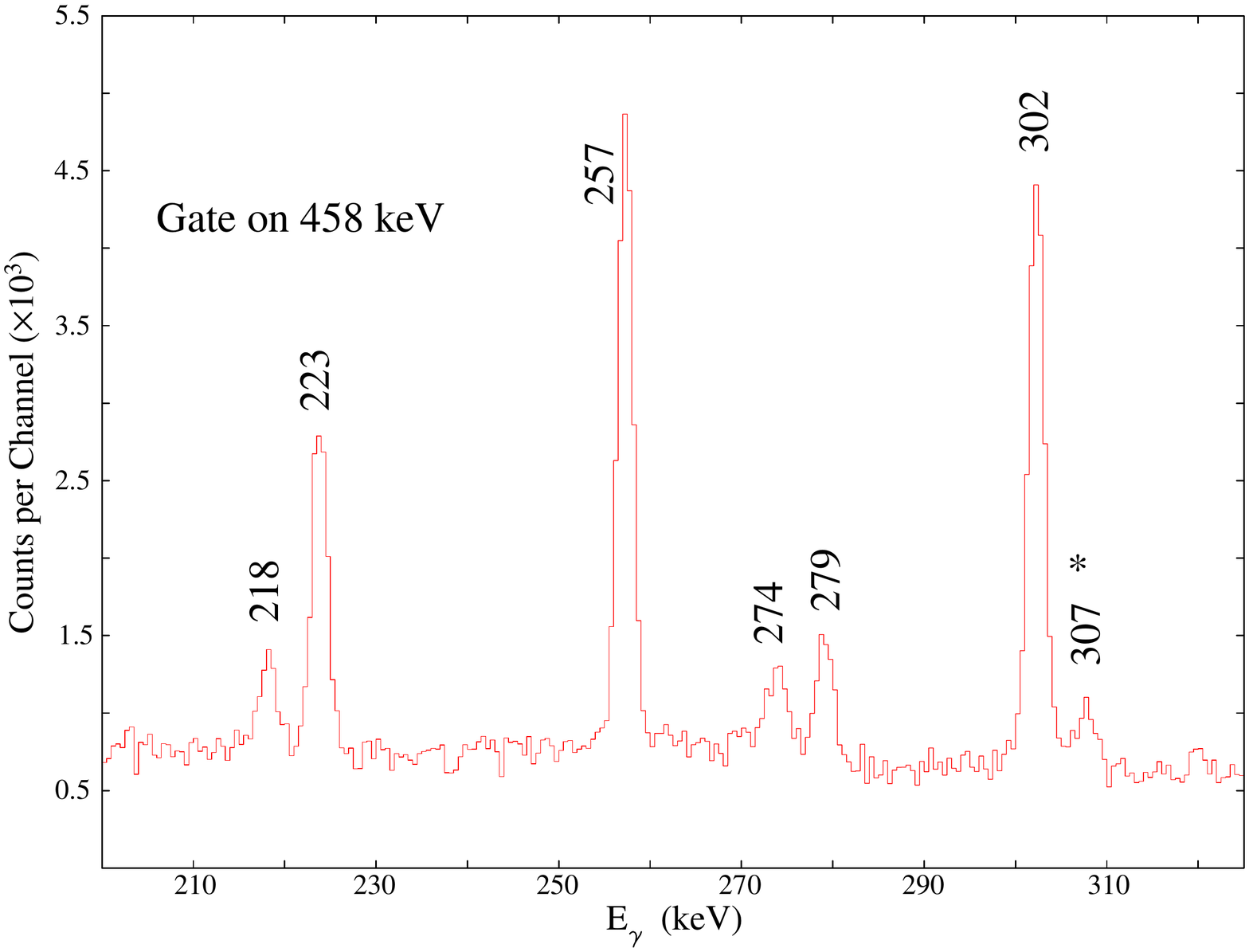}}
\subfigure[]{\label{fig:4.12b}\includegraphics[angle=0,scale=0.30,trim=0.0cm 0.0cm 0.0cm 0.0cm,clip=true]{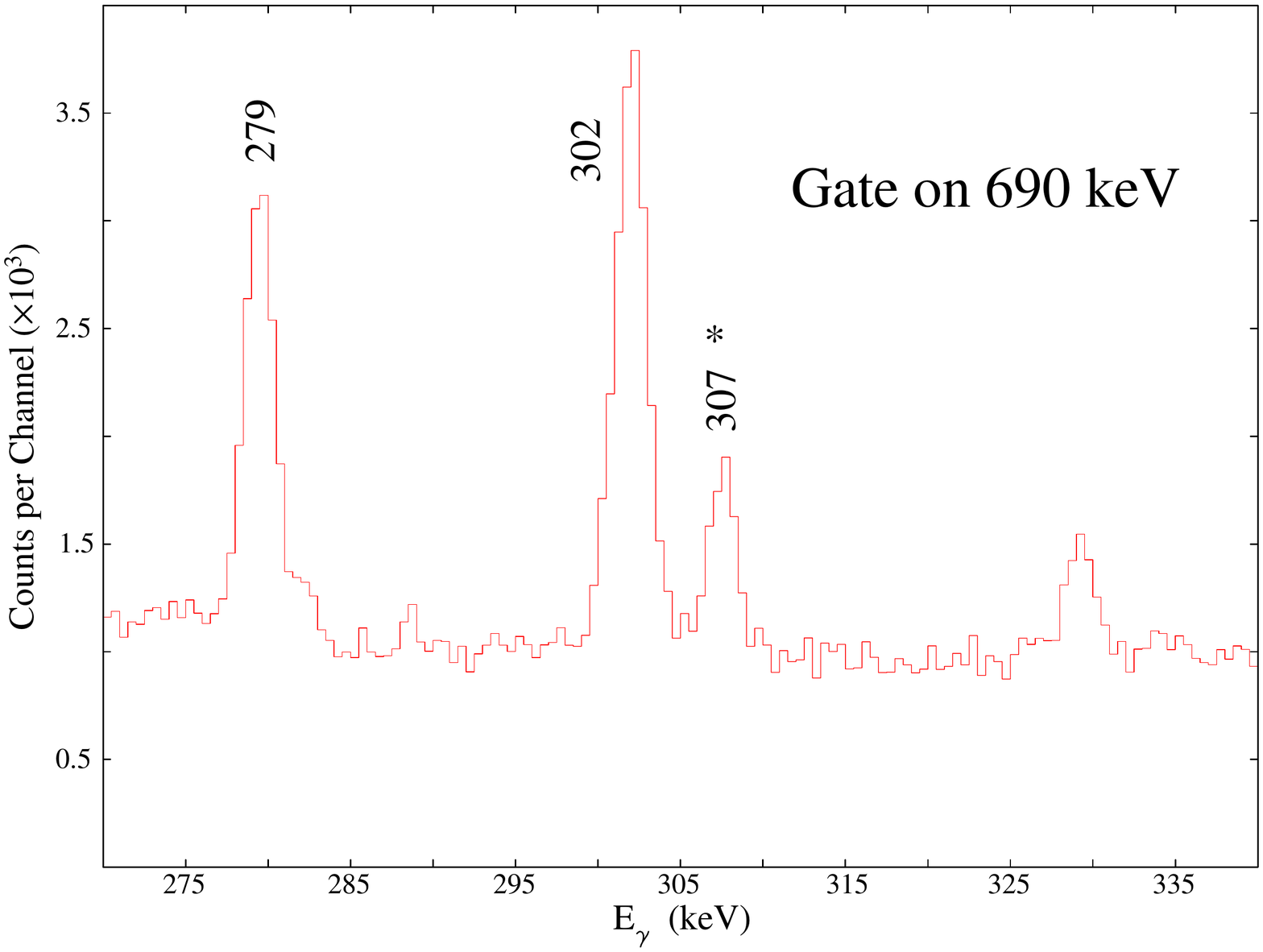}}
\caption{Representative spectra with gate on (a) 458 and (b) 690 keV transitions of $^{196}Hg$ as an evidence of 279 keV doublet transition. The new transitions,
first observed in the present study, are labeled with an asterisk (*).}
\end{figure}

Some of the existing discrepancies pertaining to the placement of $\gamma$-ray transitions in the level scheme of $^{196}$Hg were resolved following the present
measurements. For instance, the 279 keV $\gamma$-ray, de-exciting the 2344 keV, 10$^{+}$ level to 2065 keV, 9$^{-}$ level, was reported by Mehta \textit{et al.}
\cite{Meh91}, albeit with slightly different energy, as well as some of the other researchers \cite{Kro81,Gut83} but negated by Helppi \textit{et al.} \cite{Hel83}.
The transition could be confirmed in the present study and, based on $\gamma$-$\gamma$ coincidence relationships, could be distinguished from 279
(279.3) keV, 17$^{-}$ $\rightarrow$ 15$^{-}$ transition of stronger intensity in the level structure. Fig. \ref{fig:4.12a} illustrates 458 kev gate with 279
(278.8) keV (de-exciting 2344 keV state) peak but missing 279.3 keV (de-exciting 3978 keV level), whereas Fig. \ref{fig:4.12b} illustrates 690 keV gate with
279 (279.3) keV peak but the 278.8 keV transition is absent therein.\\

\begin{figure}
\subfigure[]{\label{fig:4.13a}\includegraphics[angle=0,scale=0.30,trim=0.0cm 0.0cm 0.0cm 0.0cm,clip=true]{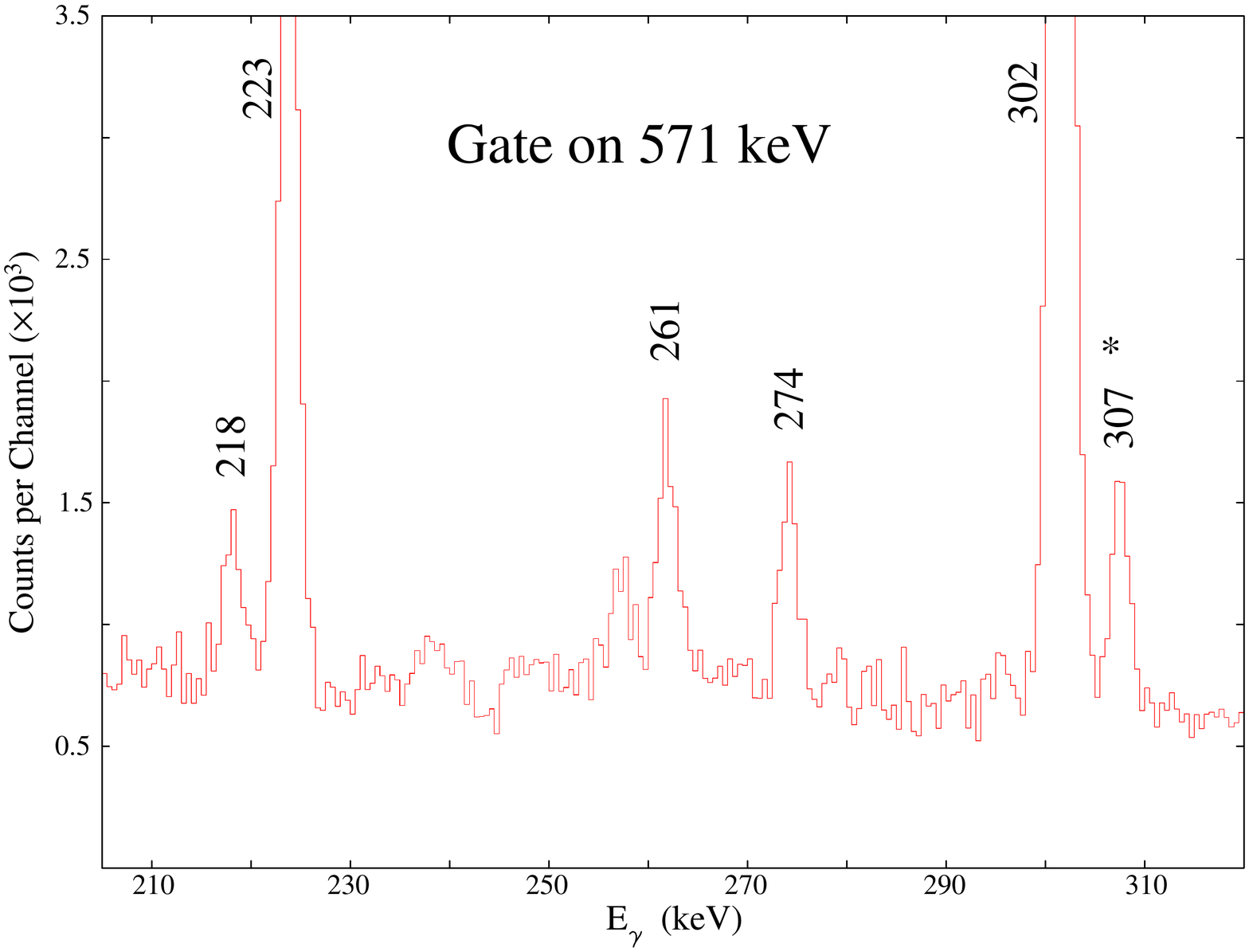}}
\subfigure[]{\label{fig:4.13b}\includegraphics[angle=0,scale=0.30,trim=0.0cm 0.0cm 0.0cm 0.0cm,clip=true]{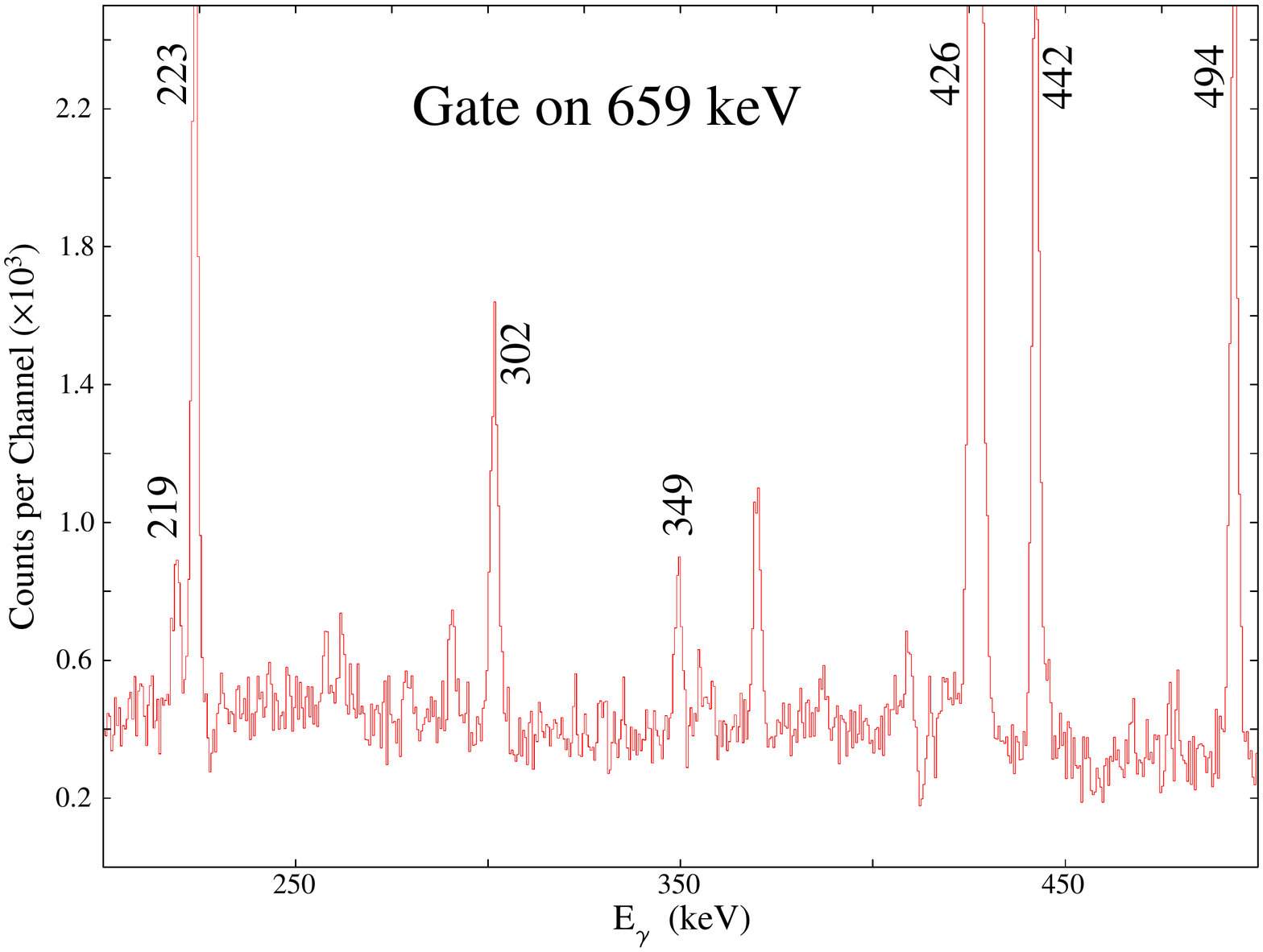}}
\caption{Representative spectra with gate on (a) 571 and (b) 659 keV transitions of $^{196}Hg$ as an evidence of 219 keV doublet transition. The new
transitions, first observed in the present study, are labeled with an asterisk (*).}
\end{figure}

Another discrepancy pertained to the observation of 218 keV transition de-exciting the 2059 keV, 7$^{-}$ level that was reported by Guttormsen \textit{et al.}
\cite{Gut83} but could not be observed by Helppi \textit{et al.} \cite{Hel83} or Mehta \textit{et al.} \cite{Meh91}. This transition was confirmed from the
present study as seen from the gate on, for instance, 571 keV transition (Fig. \ref{fig:4.13a}). The 218 keV transition was also distinguished from the close
lying 219 keV transition, first observed in the present study and de-exciting the 2284 keV, 8$^{-}$ level. The latter was confirmed in gate on 659 keV transition
(Fig. \ref{fig:4.13b}).\\

\begin{figure}
\subfigure[]{\label{fig:4.14a}\includegraphics[angle=0,scale=0.30,trim=0.0cm 0.0cm 0.0cm 0.0cm,clip=true]{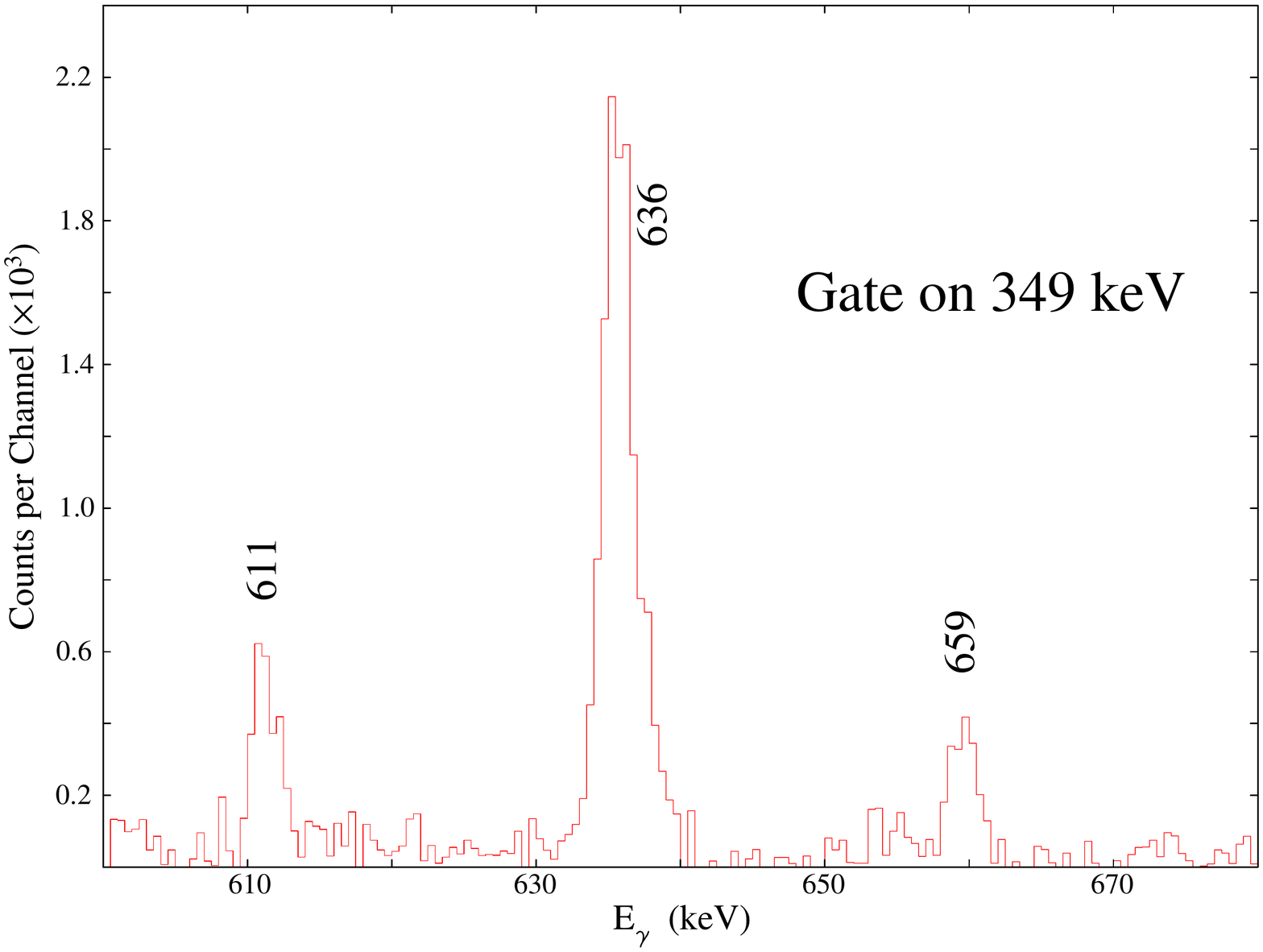}}
\subfigure[]{\label{fig:4.14b}\includegraphics[angle=0,scale=0.30,trim=0.0cm 0.0cm 0.0cm 0.0cm,clip=true]{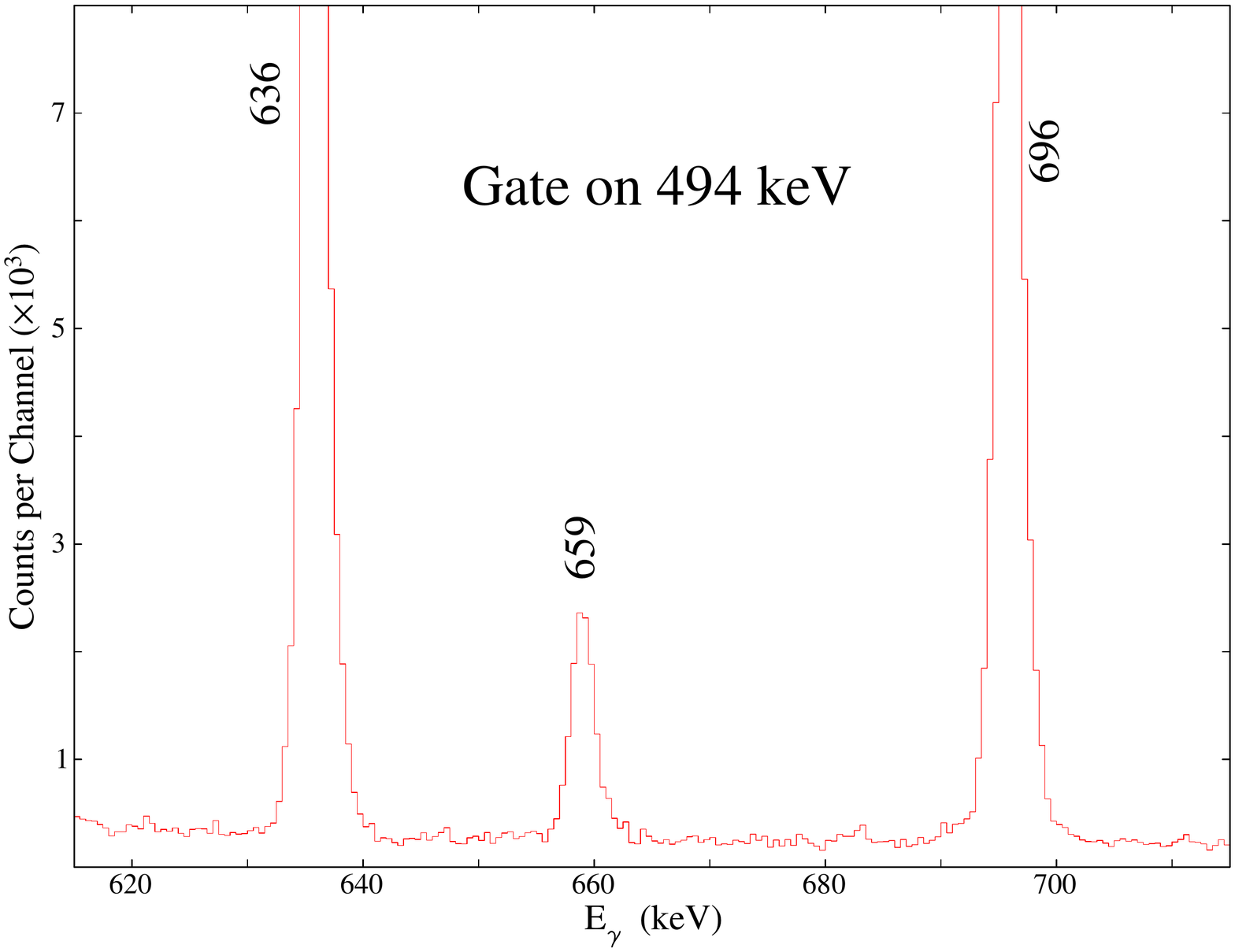}}
\caption{Representative spectra with gate on (a) 349 and (b) 494 keV transitions of $^{196}Hg$ as an evidence of 659 keV doublet transition. The new transitions,
first observed in the present study, are labeled with an asterisk (*).}
\end{figure}

There are several $\gamma$-ray transitions of very closely spaced and/or overlapping energies in the excitation scheme of $^{196}$Hg. These were placed or
confirmed through appropriate and exclusive coincidences in befitting gated spectra. An example of the exercise is the placement of 659 keV transition, one
de-exciting the 1696 keV level and the other, first observed in this study, de-exciting the 3437 keV level. The former was observed in spectra with gate on 611,
1037, 349 and 573 keV transitions while the latter in 636, 696, 84, 442 and 494 keV gates. One exclusive gated spectrum for placement of each of these 659
keV transitions is illustrated in Fig. \ref{fig:4.14a} and \ref{fig:4.14b}. Similar approach has been adopted for placements of two 714 (713.6 and 714.1) keV,
two 654 (653.5 and 653.9) keV, two 442 (441.8 and 442.1) keV, two 279 (278.8 and 279.1) keV, two 1103 (1103.2 and 1102.7) keV and two 817 (817.2 and 817.3)
keV transitions in the level structure of $^{196}$Hg.\\

\begin{longtable*}{ccccccccc}
\caption{\label{table:4.1}Details of $\gamma$-ray transitions in $^{196}$Hg nucleus observed in the present work. The new $\gamma$-ray transitions identified
from the current study are indicated by red whereas the transitions labeled in blue were observed in the previous studies but were either not placed in the
level scheme or had different placement with respect to the energy and/or $J^{\pi}$ values of the de-exciting states. $^N$ represents multipolarity adopted
from NNDC \cite{nndc}. [\,\,] indicates multipolarity assigned from the spin-parity of final and initial state without experimental measurement while (\,\,)
implies multipolarity is not assigned precisely due to large uncertainties as well as contradictory measurements.} \\
\hline
$E_{i}^{b} (keV)$ & $E_{\gamma}^{a} (keV)$ & B.R & $J_{i}^{\pi}$ & $J_{f}^{\pi}$  & $R_{ADO}$ & $\Delta_{asymm.}$ & P & Multipolarity \\
\hline
\hline
\endfirsthead

\multicolumn{9}{c}%
{{ \tablename \thetable{} -- continued from previous page}} \\
\hline
$E_{i}^{b} (keV) $ & $E_{\gamma}^{a} (keV) $ & B.R & $J_{i}^{\pi}$ & $J_{f}^{\pi}$ & $R_{ADO}$ & $\Delta_{asymm.} $ &~P~& Multipolarity \\
\hline
\endhead

\hline
\multicolumn{9}{c}{Continued in next page}\\
\hline
\endfoot
\endlastfoot
426.2$\pm$.1  & 426.2$\pm$.1                    & 1.0          & $2^{+}$  & $0^{+}$  & 1.29$\pm$.04 & 0.151$\pm$.010  &                 & E2       \\
1037.1$\pm$.1 & 610.9$\pm$.1                    & 0.78$\pm$.03 & $2^{+}$  & $2^{+}$  & 1.09$\pm$.03 & -0.045$\pm$.011 &                 & M1+E2    \\
              & 1036.6$\pm$.1                   & 0.22$\pm$.02 & $2^{+}$  & $0^{+}$  & 1.29$\pm$.11 & 0.096$\pm$.027  & 0.612$\pm$.187  & E2       \\
1062.1$\pm$.1 & 635.9$\pm$.1                    & 0.99$\pm$.01 & $4^{+}$  & $2^{+}$  & 1.27$\pm$.01 & 0.088$\pm$.004  &                 & E2       \\
              & 1061.6$\pm$.4 & 0.01$\pm$.00 & $4^{+}$  & $0^{+}$  &              &                 &                 & [E4]     \\
1390.7$\pm$.1 & 330.0$\pm$.2                    & 0.18$\pm$.02 & ($4^{-}$)& $4^{+}$  & 0.70$\pm$.07 & 0.079$\pm$.039  &                 & (E1)     \\
              & 354.8$\pm$.1                    & 0.16$\pm$.02 & ($4^{-}$)& $2^{+}$  &              &                 &                 &          \\
              & 964.5$\pm$.1                    & 0.66$\pm$.04 & ($4^{-}$)& $2^{+}$  & 1.38$\pm$.08 & -0.065$\pm$.012 &                 & (M2)     \\
1696.2$\pm$.1 & 659.1$\pm$.1                    & 1.0          &          & $2^{+}$  &              & 0.107$\pm$.018  &                 & (E)      \\
1755.3$\pm$.6 & 1329.1$\pm$.6  & 1.0          & $4^{+}$  & $2^{+}$  & 1.33$\pm$.11 & 0.108$\pm$.029  & 0.857$\pm$.253  & E2       \\
1757.7$\pm$.1 & 695.6$\pm$.1                    & 1.0          & $5^{-}$  & $4^{+}$  & 0.78$\pm$.01 & 0.051$\pm$.004  &                 & E1       \\
1775.7$\pm$.1 & 713.6$\pm$.1                    &              & $4^{+}$  & $4^{+}$  &              &                 &                 & [M1+E2]  \\
              & 738.3$\pm$.4                    &              & $4^{+}$  & $2^{+}$  &              &                 &                 & [E2]     \\
              & 1348.2$\pm$.5                   &              & $4^{+}$  & $2^{+}$  & 1.34$\pm$.13 & 0.153$\pm$.036  & 1.231$\pm$.326  & E2       \\
1785.9$\pm$.1 & 723.8$\pm$.1                    & 1.0          & $6^{+}$  & $4^{+}$  & 1.27$\pm$.01 & 0.093$\pm$.005  &                 & E2       \\
1842.2$\pm$.4 & 84.5$\pm$.4                     & 1.0          & $7^{-}$  & $5^{-}$  &              &                 &                 & $E2^{N}$ \\
1988.3$\pm$.5 & 951.2$\pm$.5                    & 1.0          &          & $2^{+}$  &              & 0.075$\pm$.034  &                 & E        \\
2012.0$\pm$.5 & 1585.8$\pm$.5                   & 1.0          &          & $2^{+}$  &              &                 &                 &          \\
2044.2$\pm$.1 & 348.5$\pm$.1                    & 0.06$\pm$.03 &          &          &              &                 &                 & E        \\
              & 653.5$\pm$.1                    & 0.68$\pm$.08 & (6)      & ($4^{-}$)& 1.34$\pm$.10 &                 &                 & (Q)      \\
              & 982.8$\pm$.1                    & 0.26$\pm$.05 & (6)      & $4^{+}$  & 0.96$\pm$.14 & -0.013$\pm$.036 &                 & (M1+E2)  \\
2059.8$\pm$.1 & 217.8$\pm$.1                    & 0.06$\pm$.01 & $7^{-}$  & $7^{-}$  & 0.98$\pm$.22 &                 &                 & D+Q      \\
              & 273.9$\pm$.1                    & 0.05$\pm$.01 & $7^{-}$  & $6^{+}$  & 0.70$\pm$.10 & 0.088$\pm$.029  & 0.268$\pm$.093  & E1       \\
              & 302.1$\pm$.1                    & 0.89$\pm$.04 & $7^{-}$  & $5^{-}$  & 1.19$\pm$.11 &                 &                 & [E2]     \\
2065.3$\pm$.5 & 223.1$\pm$.2                    & 0.97$\pm$.05 & $9^{-}$  & $7^{-}$  & 1.27$\pm$.04 & 0.163$\pm$.016  &                 & E2       \\
              & 307.2$\pm$.1   & 0.03$\pm$.01 & $9^{-}$  & $5^{-}$  &              &                 &                 & [E4]     \\
2074.3$\pm$.5 & 1648.1$\pm$.5  & 1.0          &          & $2^{+}$  &              &                 &                 &          \\
2099.2$\pm$.4 & 257.0$\pm$.1                    & 1.0          & $8^{-}$  & $7^{-}$  & 0.60$\pm$.04 & -0.037$\pm$.015 &                 & M1+E2    \\
2121.4$\pm$.5 & 1695.3$\pm$.5  & 1.0          &          & $2^{+}$  &              &                 &                 &          \\
2161.1$\pm$.5 & 1734.9$\pm$.5  & 1.0          &          & $2^{+}$  &              &                 &                 &          \\
2199.9$\pm$.5 & 1773.7$\pm$.5  & 1.0          &          & $2^{+}$  &              &                 &                 &          \\
2221.3$\pm$.1 & 445.6$\pm$.1                    & 1.0          & (6)      & $4^{+}$  & 1.20$\pm$.16 &                 &                 & (Q)      \\
2263.9$\pm$.1 & 478.0$\pm$.1                    & 1.0          & $8^{+}$  & $6^{+}$  & 1.22$\pm$.07 & 0.147$\pm$.038  & 0.555$\pm$.156  & E2       \\
2282.3$\pm$.4 & 1220.2$\pm$.4  & 1.0          &          & $4^{+}$  &              &                 &                 &          \\
2284.3$\pm$.4 & 219.1$\pm$.1   & 0.12$\pm$.03 & $8^{-}$  & $9^{-}$  & 0.91$\pm$.11 &                 &                 & (D)      \\
              & 442.1$\pm$.1  & 0.88$\pm$.06 & $8^{-}$  & $7^{-}$  & 0.65$\pm$.02 & -0.033$\pm$.015 &                 & M1+E2    \\
2340.4$\pm$.9 & 1278.3$\pm$.9  & 1.0          &          & $4^{+}$  &              &                 &                 &          \\
2344.1$\pm$.5 & 79.9$\pm$.2                     & 0.45$\pm$.12 & $10^{+}$ & $8^{+}$  &              &                 &                 & $E2^{N}$ \\
              & 278.8$\pm$.1                    & 0.55$\pm$.16 & $10^{+}$ & $9^{-}$  &              &                 &                 & [E1]     \\
2348.3$\pm$.4 & 506.1$\pm$.2                    &              & $5^{-}$  & $7^{-}$  & 1.33$\pm$.05 & 0.113$\pm$.029  & 0.439$\pm$.123  & E2       \\
              & 589.7$\pm$.1                    &              & $5^{-}$  & $5^{-}$  & 1.09$\pm$.09 &                 &                 & D+Q      \\
2360.3$\pm$.4 & 261.1$\pm$.1                    & 0.38$\pm$.10 & $9^{-}$  & $8^{-}$  & 0.73$\pm$.07 & -0.150$\pm$.087 & -0.451$\pm$.266 & M1       \\
              & 301.3$\pm$.1                    & 0.36$\pm$.08 & $9^{-}$  & $7^{-}$  & 1.37$\pm$.17 &                 &                 & Q        \\
              & 518.1$\pm$.3   & 0.26$\pm$.08 & $9^{-}$  & $7^{-}$  & 1.29$\pm$.06 & 0.129$\pm$.024  & 0.508$\pm$.110  & E2       \\
2362.3$\pm$.1 & 604.6$\pm$.1   & 1.0          &          & $5^{-}$  &              & -0.039$\pm$.014 &                 & M        \\
2441.0$\pm$.5 & 96.9$\pm$.2                     & 1.0          & $12^{+}$ & $10^{+}$ &              &                 &                 & $E2^{N}$ \\
2493.9$\pm$.1 & 1103.2$\pm$.1  & 1.0          &          &          &              &                 &                 &          \\
2495.6$\pm$.6 & 1433.5$\pm$.4                   & 1.0          &          & $4^{+}$  &              & 0.123$\pm$.042  &                 & E        \\
2518.2$\pm$.4 & 419.0$\pm$.1   & 1.0          & $9^{(+)}$& $8^{-}$  & 0.70$\pm$.09 & 0.016$\pm$.041  &                 & (E)1     \\
2555.4$\pm$.4 & 456.2$\pm$.1                    & 0.69$\pm$.06 & $10^{-}$ & $8^{-}$  & 1.33$\pm$.07 & 0.125$\pm$.018  & 0.461$\pm$.084  & E2       \\
              & 489.5$\pm$.1                    & 0.31$\pm$.04 & $10^{-}$ & $9^{-}$  & 1.08$\pm$.11 & -0.080$\pm$.024 &                 & M1+E2    \\
2609.5$\pm$.5 & 1547.4$\pm$.5  & 1.0          &          & $4^{+}$  &              & 0.038$\pm$.056  &                 & E        \\
2617.6$\pm$.1 & 573.4$\pm$.1   & 1.0          & (7)      & (6)      & 1.05$\pm$.15 & 0.121$\pm$.042  &                 & (E1+M2)  \\
2621.8$\pm$.5 & 556.5$\pm$.1                    & 1.0          & $11^{-}$ & $9^{-}$  & 1.31$\pm$.02 & 0.113$\pm$.019  & 0.463$\pm$.094  & E2       \\
2638.8$\pm$.4 & 539.6$\pm$.1   & 1.0          &          &          &              &                 &                 &          \\
2721.6$\pm$.5 & 1659.5$\pm$.5  & 1.0          &          & $4^{+}$  &              &                 &                 &          \\
2768.0$\pm$.5 & 1705.9$\pm$.5  & 1.0          &          & $4^{+}$  &              &                 &                 &          \\
2776.7$\pm$1.0& 1714.6$\pm$1.0 & 1.0          &          & $4^{+}$  &              &                 &                 &          \\
2778.6$\pm$.1 & 494.3$\pm$.2   & 0.42$\pm$.08 & $10^{-}$ & $8^{-}$  & 1.31$\pm$.07 & 0.126$\pm$.020  & 0.484$\pm$.094  & E2       \\
              & 712.5$\pm$.5   & 0.15$\pm$.03 & $10^{-}$ & $9^{-}$  &              &                 &                 & [M1+E2]  \\
              & 734.4$\pm$.1                    & 0.43$\pm$.05 & $10^{-}$ & (6)      & 1.21$\pm$.15 & 0.142$\pm$.038  &                 & (E2)     \\
2802.0$\pm$.5 & 441.8$\pm$.5  & 0.29$\pm$.07 & $8^{+}$  & $9^{-}$  &              &                 &                 & [E1]     \\
              & 457.9$\pm$.2   & 0.71$\pm$.11 & $8^{+}$  & $10^{+}$ & 1.30$\pm$.08 & 0.101$\pm$.020  & 0.373$\pm$.085  & E2       \\
2841.3$\pm$.5 & 497.2$\pm$.1   &              &          & $10^{+}$ &              & -0.030$\pm$.058 &                 & (M+E)    \\
              & 576.4$\pm$.1   &              &          & $8^{+}$  &              & 0.156$\pm$.051  &                 & E        \\
2845.9$\pm$.5 & 404.9$\pm$.1                    & 1.0          & $14^{+}$ & $12^{+}$ & 1.27$\pm$.05 & 0.104$\pm$.013  &                 & E2       \\
2917.8$\pm$.1 & 653.9$\pm$.1   & 1.0          &          & $6^{+}$  &              & 0.031$\pm$.021  &                 & E        \\
2924.5$\pm$.8 & 1138.6$\pm$.8  & 1.0          & 7        & $6^{+}$  & 1.09$\pm$.22 &                 &                 & D+Q      \\
2931.1$\pm$.4 & 570.8$\pm$.1                    & 1.0          & $11^{-}$ & $9^{-}$  & 1.28$\pm$.08 & 0.124$\pm$.027  & 0.516$\pm$.126  & E2       \\
2940.7$\pm$.5 & 875.4$\pm$.1   & 1.0          & $10^{+}$ & $9^{-}$  & 0.84$\pm$.09 & 0.063$\pm$.025  & 0.350$\pm$.145  & E1       \\
2978.0$\pm$.2 & 714.1$\pm$.2  & 1.0          & $10^{+}$ & $8^{+}$  & 1.24$\pm$.08 & 0.112$\pm$.034  & 0.537$\pm$.174  & E2       \\
2993.2$\pm$.3 & 729.3$\pm$.3   & 1.0          &          & $8^{+}$  &              & -0.060$\pm$.032 &                 & M        \\
3101.5$\pm$.4 & 817.2$\pm$.1   & 1.0          &          &          &              &                 &                 &          \\
3135.6$\pm$.4 & 775.3$\pm$.1                    & 1.0          &          &          &              &                 &                 &          \\
3201.6$\pm$.5 & 760.6$\pm$.1                    & 1.0          &          & $12^{+}$ &              & 0.083$\pm$.022  &                 & E        \\
3238.4$\pm$.4 & 683.0$\pm$.1                    &              & $12^{-}$ & $10^{-}$ & 1.29$\pm$.11 & 0.098$\pm$.025  & 0.456$\pm$.127  & E2       \\
              & 1173.4$\pm$.7  &              & $12^{-}$ & $9^{-}$  & 0.71$\pm$.11 & 0.048$\pm$.033  &                 &          \\
3312.1$\pm$.5 & 690.3$\pm$.1                    & 1.0          & $13^{-}$ & $11^{-}$ & 1.26$\pm$.06 & 0.125$\pm$.029  & 0.586$\pm$.151  & E2       \\
3368.2$\pm$.5 & 566.2$\pm$.1   & 1.0          &          &          &              &                 &                 &          \\
3370.0$\pm$.5 & 524.1$\pm$.2   & 1.0          &          & $14^{+}$ &              & -0.020$\pm$.047 &                 & M        \\
3373.1$\pm$.8 & 571.1$\pm$.6  & 1.0          &          &          &              &                 &                 &          \\
3404.9$\pm$.5 & 559.0$\pm$.1                    & 1.0          & $16^{+}$ & $14^{+}$ & 1.25$\pm$.11 & 0.132$\pm$.025  & 0.542$\pm$.120  & E2       \\
3437.4$\pm$.1 & 658.8$\pm$.1   & 1.0          & $12^{-}$ & $10^{-}$ & 1.29$\pm$.11 & 0.135$\pm$.036  & 0.613$\pm$.178  & E2       \\
3439.1$\pm$.5 & 817.3$\pm$.1   & 1.0          & 13       & $11^{-}$ & 1.31$\pm$.28 &                 &                 & Q        \\
3476.7$\pm$.6 & 630.8$\pm$.4   & 1.0          & 16       & $14^{+}$ & 1.26$\pm$.09 &                 &                 & Q        \\
3509.5$\pm$.5 & 663.6$\pm$.1                    & 1.0          & $16^{+}$ & $14^{+}$ & 1.28$\pm$.12 & 0.089$\pm$.026  & 0.406$\pm$.127  & E2       \\
3532.1$\pm$.1 & 753.5$\pm$.1   &              & $12^{-}$ & $10^{-}$ &              &                 &                 & [E2]     \\
              & 909.4$\pm$.5   &              & $12^{-}$ & $11^{-}$ & 0.71$\pm$.07 & -0.050$\pm$.026 & -0.286$\pm$.153 & M1       \\
3539.0$\pm$.5 & 693.1$\pm$.2  & 1.0          & $15^{-}$ & $14^{+}$ & 0.85$\pm$.04 & 0.066$\pm$.031  & 0.310$\pm$.150  & E1       \\
3592.8$\pm$.5 & 971.0$\pm$.1   &              & 13       & $11^{-}$ & 1.30$\pm$.15 &                 &                 & Q        \\
              & 662.3$\pm$.6   &              & 13       & $11^{-}$ &              &                 &                 &          \\
3623.1$\pm$.5 & 1359.2$\pm$.5  & 1.0          &          & $8^{+}$  &              & 0.130$\pm$.036  &                 & E        \\
3633.1$\pm$.5 & 1369.2$\pm$.5  & 1.0          &          & $8^{+}$  &              &                 &                 &          \\
3649.3$\pm$.5 & 1385.4$\pm$.5  & 1.0          &          & $8^{+}$  &              &                 &                 &          \\
3686.2$\pm$.5 & 840.3$\pm$.1                    & 1.0          &          & $14^{+}$ &              & 0.106$\pm$.026  &                 & E        \\
3698.9$\pm$.5 & 386.8$\pm$.1                    & 0.78$\pm$.13 & $15^{-}$ & $13^{-}$ & 1.27$\pm$.08 & 0.140$\pm$.030  & 0.480$\pm$.116  & E2       \\
              & 853.3$\pm$.2                    & 0.22$\pm$.05 & $15^{-}$ & $14^{+}$ &              &                 &                 & [E1]     \\
3724.6$\pm$.6 & 793.6$\pm$.1   &              & $13^{-}$ & $11^{-}$ & 1.32$\pm$.10 &                 &                 & E2       \\
              & 1102.7$\pm$.3  &              & $13^{-}$ & $11^{-}$ &              & 0.073$\pm$.052  & 0.490$\pm$.354  & E2       \\
3793.5$\pm$.5 & 947.6$\pm$.1                    & 1.0          & $15^{+}$ & $14^{+}$ & 0.68$\pm$.09 & -0.091$\pm$.044 &                 & M1+E2    \\
3978.2$\pm$.5 & 279.3$\pm$.1                    & 1.0          & 17       & $15^{-}$ & 1.30$\pm$.17 &                 &                 & Q        \\
4110.9$\pm$.5 & 798.8$\pm$.1                    & 1.0          & $15^{-}$ & $13^{-}$ & 1.25$\pm$.17 & 0.093$\pm$.032  & 0.483$\pm$.175  & E2       \\
4323.1$\pm$.5 & 813.6$\pm$.2                    & 1.0          & $18^{+}$ & $16^{+}$ &              &                 &                 & $E2^{N}$ \\
\hline
\hline
\bigskip
\end{longtable*}

\begin{figure}
\includegraphics[angle=0,scale=0.45,trim=0.0cm 0.0cm 1.5cm 1.0cm,clip=true]{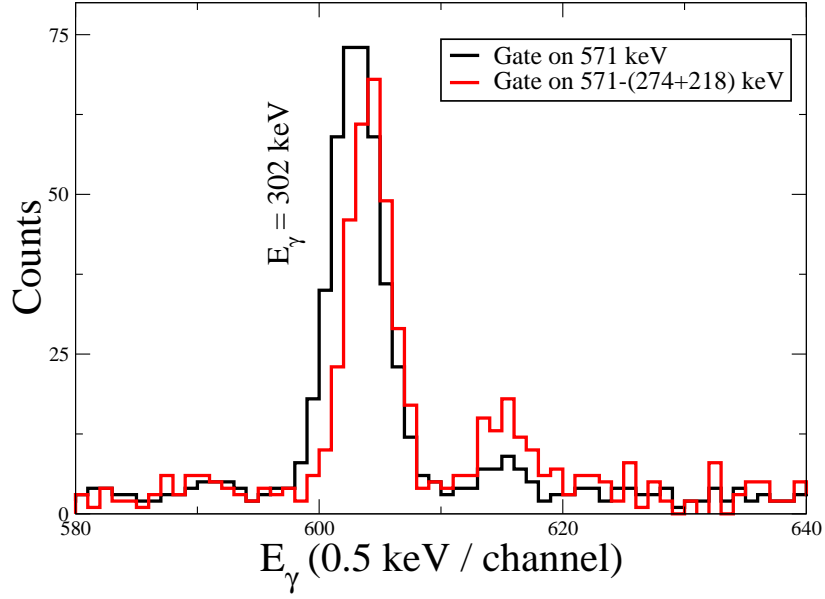}
\caption{\label{fig:4.15}Comparison of intensity difference in 302 keV transition for gate on 571 keV, a sum gate on 274 and 218 keV transitions subtracted
from that on 571 keV.}
\end{figure}

Two additional negative parity sequences that were proposed by Helppi \textit{et al.} \cite{Hel83} and Mehta \textit{et al.} \cite{Meh91}, have been confirmed
and partially extended in the present study. One of these is the 7$^{-}$ $\leftarrow$ 9$^{-}$ $\leftarrow$ 11$^{-}$ $\leftarrow$ 13$^{-}$ sequence connected
by 301, 571, 794 keV transitions of which the top most 794 keV transition was first observed in this investigation. The band head of this sequence, at 2060
keV, was tentatively assigned spin-parity 6$^{-}$ by Mehta \textit{et al.} and previously by Helppi \textit{et al.}, as well. The spin-parity assignments for
the other levels of this sequence were missing in either of these earlier studies. Based on the present R$_{ADO}$ and polarization measurement of the 274 keV
$\gamma$-ray transition, de-exciting the 2060 keV state, the spin-parity of the latter was modified to 7$^{-}$. The 302 keV transition, de-exciting the same
2060 keV level and more intense than the 274 keV one (Table \ref{table:4.1}), could not be directly used for spin-parity assignment of the state owing to the
overlap between this 302 keV (5$^{-}$ $\leftarrow$ 7$^{-}$) transition and the coincident 301 keV (7$^{-}$ $\leftarrow$ 9$^{-}$) transition de-exciting the
2360 keV level. There was no gating transition identified exclusively for determining the R$_{ADO}$ and/or polarization value of the former. However, a sum
of gates on the 274 keV and the 218 keV transitions when subtracted from that corresponding to a gate 571 keV (Fig. \ref{fig:4.15}) transition is expected to
eliminate the contribution of 301 keV (7$^{-}$ $\leftarrow$ 9$^{-}$) transition from the 302 keV (5$^{-}$ $\leftarrow$ 7$^{-}$) peak. The same was used for
determining the R$_{ADO}$ value of the latter and the result upheld the 7$^{-}$ assignment to the 2060 keV state. The present measurements were also extended
to other transitions in the sequence and corresponding assignments were made therefrom. The other negative parity sequence 8$^{-}$ $\leftarrow$ 10$^{-}$
$\leftarrow$ 12$^{-}$, that was identified by Helppi \textit{et al.} \cite{Hel83} and later by Mehta \textit{et al.} \cite{Meh91}, was also confirmed in the
current investigation. The spin-parity assignments therein could also be validated from the R$_{ADO}$ and polarization measurements of the intraband transitions
carried out in the present work.\\

\begin{figure}
\includegraphics[angle=0,scale=0.30,trim=0.0cm 0.0cm 0.0cm 0.0cm,clip=true]{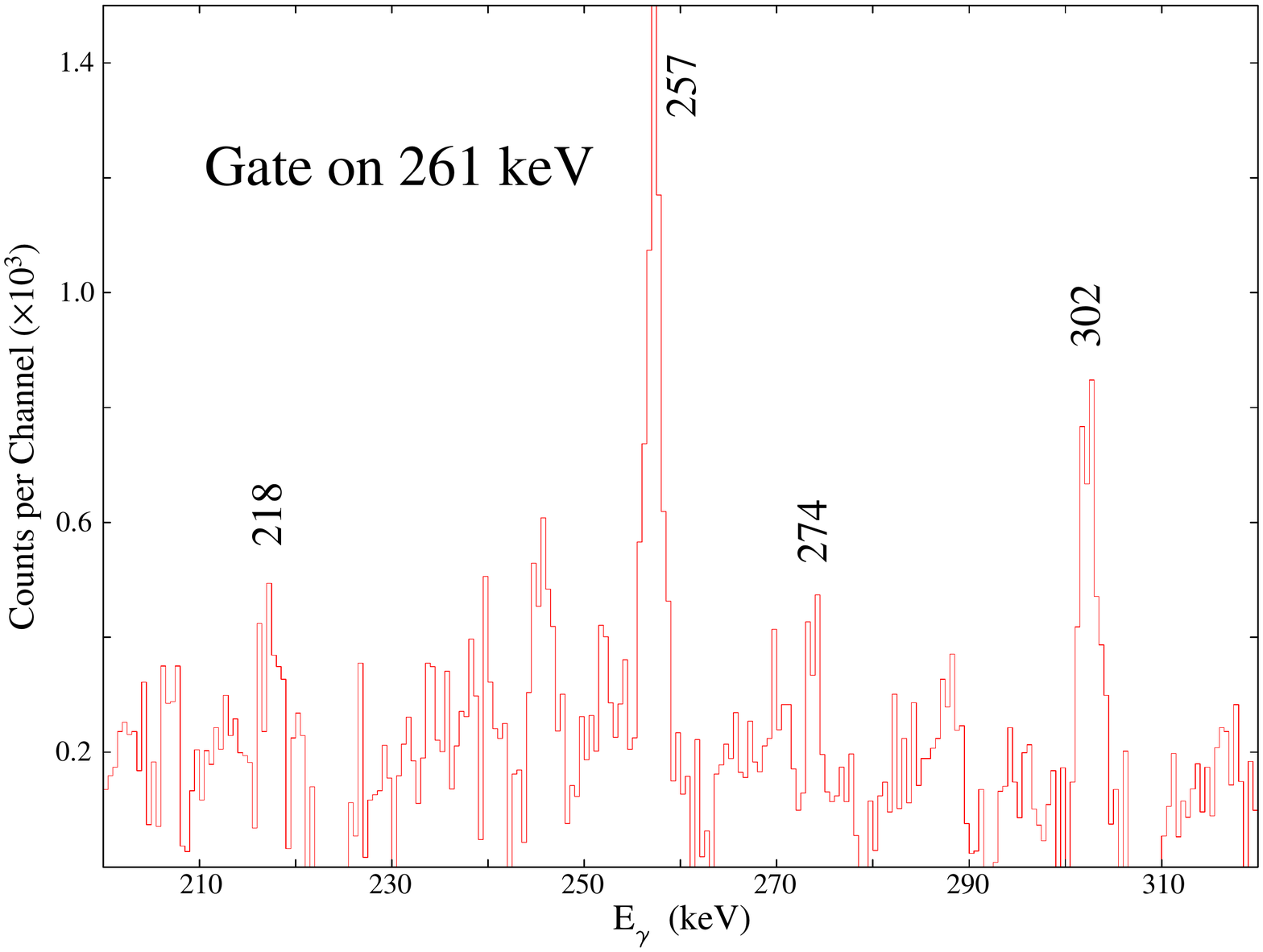}
\includegraphics[angle=0,scale=0.30,trim=0.0cm 0.0cm 0.0cm 0.0cm,clip=true]{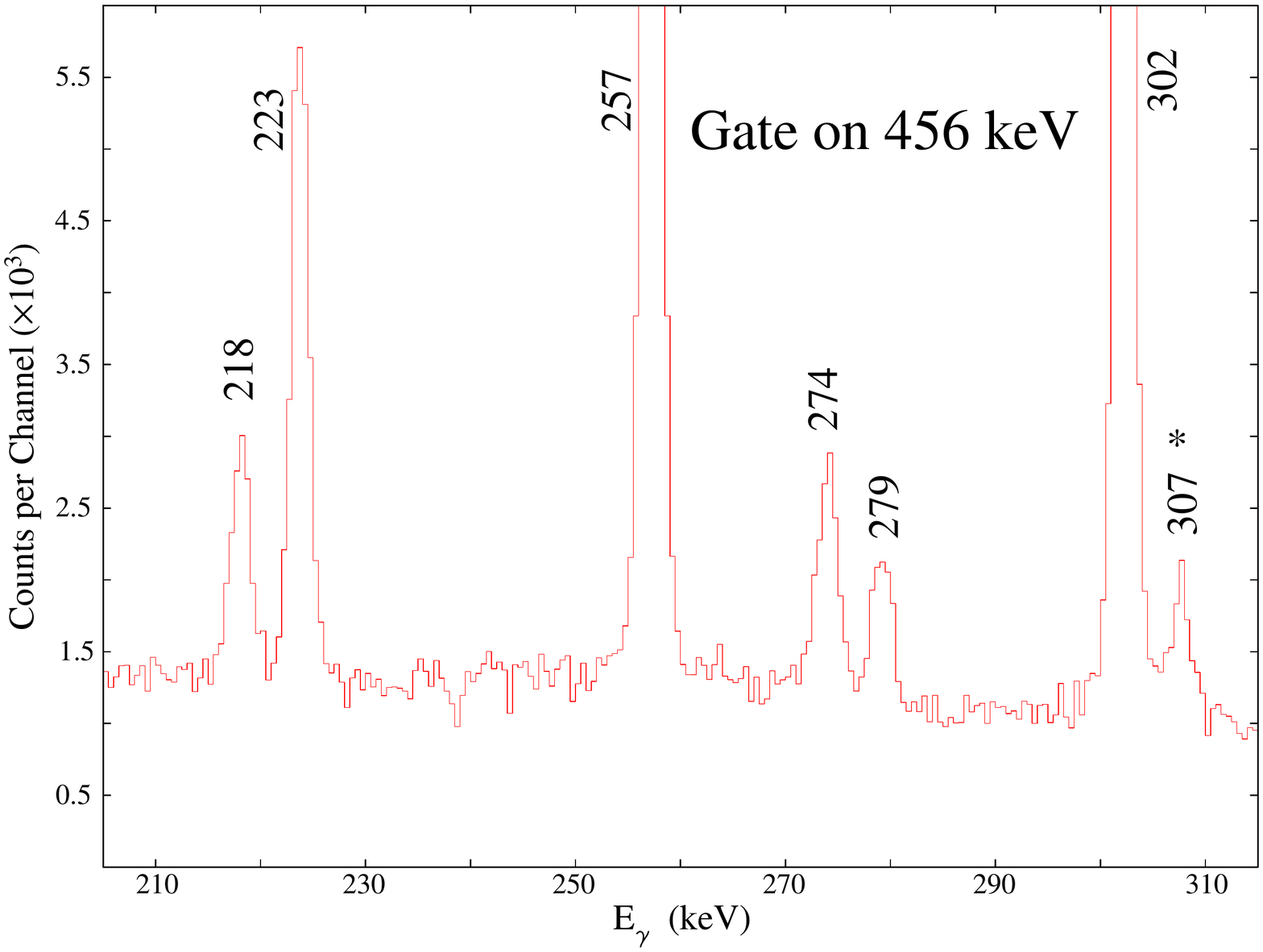}
\caption{\label{fig:4.16}Representative spectra with gate on (a) 261 and (b) 456 keV transitions of $^{196}Hg$ relates the existence of 39 keV transition. The
new transitions, first observed in the present study, are labeled with an asterisk (*) }
\end{figure}

Interestingly, the three $\gamma$-ray transitions, de-exciting the 2060 keV level, 302, 274 and 218 keV, are observed in coincidence with 261, 456 and 683 keV
transitions, as illustrated in Fig. \ref{fig:4.16}, spectrum corresponding to a gate on the 261 and 456 keV. The same coincidences had been previously observed by
Bernards \textit{et al.} \cite{Ber10} and indicate a 39 keV transition connecting the 2099 keV level to the 2060 keV state, albeit the same was not observed in
the present experiment. This was owing to the energy thresholds set on the detector channels, for the current measurement, and could have been facilitated by a
LEPS detector not used during this experiment.\\

A new sequence of states 2284 (8$^{-}$) $\leftarrow$ 2779 (10$^{-}$) $\leftarrow$ 3437 (12$^{-}$) was observed for the first time in the present investigation,
albeit the 2779 keV level had been previously reported by Bernards \textit{et al.} \cite{Ber10}, to be de-exciting through 734 keV transition, with tentative
spin-parity assignments. The R$_{ADO}$ and polarization were determined for the 494 keV and 659 keV transitions of the sequence, first observed in the present
work, and spin-parity assignments were made to the respective states.\\

The remaining findings in the level structure of $^{196}$Hg from the present efforts can be summarized as, (i) high energy $\gamma$-ray transitions in the energy
range 1.5 MeV to 1.7 MeV feeding the first (2$^{+}$) or the second (4$^{+}$) excited state of the nucleus and (ii) $\gamma$-ray transitions feeding higher excited
states in the aforementioned sequences or connecting them. The former might be of single particle origin while the latter may plausibly represent intriguing
structural evolution underlying the sequences/bands. High energy $\gamma$-ray transitions of close lying energies ($\sim$ 1.3 MeV) were also observed to feed
the 2264 keV level of ground state positive parity band. These are maiden observations in the present study.\\

\section{Summary}

Spectroscopic investigation of the $^{196}$Hg (Z=80, N=116) was carried out using the INGA setup at VECC supported by the digitizer based pulse processing and
data acquisition system \cite{Das18}. Analysis of the acquired data also led to the observation of new $\gamma$-ray transitions and levels as well as to spin-parity
assignments thereof. The latter were either previously absent or had been tentatively identified. Theoretical interpretation of the level structure obtained from
these efforts can be pursued with appropriate model calculations.

\bibliography{196Hg_arxiv_v3.bib}

\begin{thebibliography}{15}
\expandafter\ifx\csname natexlab\endcsname\relax\def\natexlab#1{#1}\fi
\expandafter\ifx\csname bibnamefont\endcsname\relax
  \def\bibnamefont#1{#1}\fi
\expandafter\ifx\csname bibfnamefont\endcsname\relax
  \def\bibfnamefont#1{#1}\fi
\expandafter\ifx\csname citenamefont\endcsname\relax
  \def\citenamefont#1{#1}\fi
\expandafter\ifx\csname url\endcsname\relax
  \def\url#1{\texttt{#1}}\fi
\expandafter\ifx\csname urlprefix\endcsname\relax\def\urlprefix{URL }\fi
\providecommand{\bibinfo}[2]{#2}
\providecommand{\eprint}[2][]{\url{#2}}

\bibitem[{\citenamefont{Lieder et~al.}(1975)\citenamefont{Lieder, Beuscher,
  Davidson, Neskakis, and Mayer-Boricke}}]{Lie75}
\bibinfo{author}{\bibfnamefont{R.~M.} \bibnamefont{Lieder}},
  \bibinfo{author}{\bibfnamefont{H.}~\bibnamefont{Beuscher}},
  \bibinfo{author}{\bibfnamefont{W.~F.} \bibnamefont{Davidson}},
  \bibinfo{author}{\bibfnamefont{A.}~\bibnamefont{Neskakis}}, \bibnamefont{and}
  \bibinfo{author}{\bibfnamefont{C.}~\bibnamefont{Mayer-Boricke}},
  \bibinfo{journal}{Nucl. \ Phy. \ A} \textbf{\bibinfo{volume}{248}},
  \bibinfo{pages}{317} (\bibinfo{year}{1975}).

\bibitem[{\citenamefont{Gunther et~al.}(1977)\citenamefont{Gunther, Hubel,
  Kleinrahm, Mertin, Richter, Schneider, and Tischler}}]{Gun77}
\bibinfo{author}{\bibfnamefont{C.}~\bibnamefont{Gunther}},
  \bibinfo{author}{\bibfnamefont{H.}~\bibnamefont{Hubel}},
  \bibinfo{author}{\bibfnamefont{A.}~\bibnamefont{Kleinrahm}},
  \bibinfo{author}{\bibfnamefont{D.}~\bibnamefont{Mertin}},
  \bibinfo{author}{\bibfnamefont{B.}~\bibnamefont{Richter}},
  \bibinfo{author}{\bibfnamefont{W.~D.} \bibnamefont{Schneider}},
  \bibnamefont{and} \bibinfo{author}{\bibfnamefont{R.}~\bibnamefont{Tischler}},
  \bibinfo{journal}{Phys. \ Rev.\ C} \textbf{\bibinfo{volume}{15}},
  \bibinfo{pages}{1298} (\bibinfo{year}{1977}).

\bibitem[{\citenamefont{Proetel et~al.}(1974)\citenamefont{Proetel, Diamond,
  and Stephens}}]{Pro74}
\bibinfo{author}{\bibfnamefont{D.}~\bibnamefont{Proetel}},
  \bibinfo{author}{\bibfnamefont{R.~M.} \bibnamefont{Diamond}},
  \bibnamefont{and} \bibinfo{author}{\bibfnamefont{F.~S.}
  \bibnamefont{Stephens}}, \bibinfo{journal}{Nucl. \ Phy. \ A}
  \textbf{\bibinfo{volume}{231}}, \bibinfo{pages}{301} (\bibinfo{year}{1974}).

\bibitem[{\citenamefont{Cederwall et~al.}(1993)\citenamefont{Cederwall,
  Deleplanque, Azaiez, Diamond, Fallon, Korten, Lee, Macchiavelli, Oliveira,
  Stephens et~al.}}]{Ced93}
\bibinfo{author}{\bibfnamefont{B.}~\bibnamefont{Cederwall}},
  \bibinfo{author}{\bibfnamefont{M.~A.} \bibnamefont{Deleplanque}},
  \bibinfo{author}{\bibfnamefont{F.}~\bibnamefont{Azaiez}},
  \bibinfo{author}{\bibfnamefont{R.~M.} \bibnamefont{Diamond}},
  \bibinfo{author}{\bibfnamefont{P.}~\bibnamefont{Fallon}},
  \bibinfo{author}{\bibfnamefont{W.}~\bibnamefont{Korten}},
  \bibinfo{author}{\bibfnamefont{L.~Y.} \bibnamefont{Lee}},
  \bibinfo{author}{\bibfnamefont{A.~O.} \bibnamefont{Macchiavelli}},
  \bibinfo{author}{\bibfnamefont{J.~R.~B.} \bibnamefont{Oliveira}},
  \bibinfo{author}{\bibfnamefont{F.~S.} \bibnamefont{Stephens}},
  \bibnamefont{et~al.}, \bibinfo{journal}{Phys. \ Rev.\ C}
  \textbf{\bibinfo{volume}{47}}, \bibinfo{pages}{6} (\bibinfo{year}{1993}).

\bibitem[{\citenamefont{Helppi et~al.}(1983)\citenamefont{Helppi, Saha, Daly,
  Faber, and Khoo}}]{Hel83}
\bibinfo{author}{\bibfnamefont{H.}~\bibnamefont{Helppi}},
  \bibinfo{author}{\bibfnamefont{S.~K.} \bibnamefont{Saha}},
  \bibinfo{author}{\bibfnamefont{P.~J.} \bibnamefont{Daly}},
  \bibinfo{author}{\bibfnamefont{S.~R.} \bibnamefont{Faber}}, \bibnamefont{and}
  \bibinfo{author}{\bibfnamefont{T.~L.} \bibnamefont{Khoo}},
  \bibinfo{journal}{Phys. \ Rev.\ C} \textbf{\bibinfo{volume}{28}},
  \bibinfo{pages}{3} (\bibinfo{year}{1983}).

\bibitem[{\citenamefont{Mehta et~al.}(1991)\citenamefont{Mehta, Agarwal, Blume,
  Heppner, Hubel, Murzel, Theine, and Gast}}]{Meh91}
\bibinfo{author}{\bibfnamefont{D.}~\bibnamefont{Mehta}},
  \bibinfo{author}{\bibfnamefont{Y.~K.} \bibnamefont{Agarwal}},
  \bibinfo{author}{\bibfnamefont{K.~P.} \bibnamefont{Blume}},
  \bibinfo{author}{\bibfnamefont{S.}~\bibnamefont{Heppner}},
  \bibinfo{author}{\bibfnamefont{H.}~\bibnamefont{Hubel}},
  \bibinfo{author}{\bibfnamefont{M.}~\bibnamefont{Murzel}},
  \bibinfo{author}{\bibfnamefont{K.}~\bibnamefont{Theine}}, \bibnamefont{and}
  \bibinfo{author}{\bibfnamefont{W.}~\bibnamefont{Gast}}, \bibinfo{journal}{Z.
  \ Phys.\ A} \textbf{\bibinfo{volume}{339}}, \bibinfo{pages}{317}
  (\bibinfo{year}{1991}).

\bibitem[{\citenamefont{Kroth et~al.}(1981)\citenamefont{Kroth, Hardt,
  Guttormsen, Mikus, Recht, Vilter, Hubel, and Gunther}}]{Kro81}
\bibinfo{author}{\bibfnamefont{R.}~\bibnamefont{Kroth}},
  \bibinfo{author}{\bibfnamefont{K.}~\bibnamefont{Hardt}},
  \bibinfo{author}{\bibfnamefont{M.}~\bibnamefont{Guttormsen}},
  \bibinfo{author}{\bibfnamefont{G.}~\bibnamefont{Mikus}},
  \bibinfo{author}{\bibfnamefont{J.}~\bibnamefont{Recht}},
  \bibinfo{author}{\bibfnamefont{W.}~\bibnamefont{Vilter}},
  \bibinfo{author}{\bibfnamefont{H.}~\bibnamefont{Hubel}}, \bibnamefont{and}
  \bibinfo{author}{\bibfnamefont{C.}~\bibnamefont{Gunther}},
  \bibinfo{journal}{Phys. \ Lett.\ B} \textbf{\bibinfo{volume}{99}},
  \bibinfo{pages}{209} (\bibinfo{year}{1981}).

\bibitem[{\citenamefont{Bernards et~al.}(2010)\citenamefont{Bernards, Heinze,
  Jolie, Albers, Fransen, and Radeck}}]{Ber10}
\bibinfo{author}{\bibfnamefont{C.}~\bibnamefont{Bernards}},
  \bibinfo{author}{\bibfnamefont{S.}~\bibnamefont{Heinze}},
  \bibinfo{author}{\bibfnamefont{J.}~\bibnamefont{Jolie}},
  \bibinfo{author}{\bibfnamefont{M.}~\bibnamefont{Albers}},
  \bibinfo{author}{\bibfnamefont{C.}~\bibnamefont{Fransen}}, \bibnamefont{and}
  \bibinfo{author}{\bibfnamefont{D.}~\bibnamefont{Radeck}},
  \bibinfo{journal}{Phys. \ Rev.\ C} \textbf{\bibinfo{volume}{81}},
  \bibinfo{pages}{024312} (\bibinfo{year}{2010}).

\bibitem[{\citenamefont{Gavron}(1980)}]{Gav80}
\bibinfo{author}{\bibfnamefont{A.}~\bibnamefont{Gavron}},
  \bibinfo{journal}{Phys. \ Rev. \ C} \textbf{\bibinfo{volume}{21}},
  \bibinfo{pages}{230} (\bibinfo{year}{1980}).

\bibitem[{\citenamefont{Das et~al.}(2018)\citenamefont{Das, Samanta, Banik,
  Bhattacharjee, Basu, Raut, Ghugre, Sinha, Bhattacharya, Imran
  et~al.}}]{Das18}
\bibinfo{author}{\bibfnamefont{S.}~\bibnamefont{Das}},
  \bibinfo{author}{\bibfnamefont{S.}~\bibnamefont{Samanta}},
  \bibinfo{author}{\bibfnamefont{R.}~\bibnamefont{Banik}},
  \bibinfo{author}{\bibfnamefont{R.}~\bibnamefont{Bhattacharjee}},
  \bibinfo{author}{\bibfnamefont{K.}~\bibnamefont{Basu}},
  \bibinfo{author}{\bibfnamefont{R.}~\bibnamefont{Raut}},
  \bibinfo{author}{\bibfnamefont{S.~S.} \bibnamefont{Ghugre}},
  \bibinfo{author}{\bibfnamefont{A.~K.} \bibnamefont{Sinha}},
  \bibinfo{author}{\bibfnamefont{S.}~\bibnamefont{Bhattacharya}},
  \bibinfo{author}{\bibfnamefont{S.}~\bibnamefont{Imran}},
  \bibnamefont{et~al.}, \bibinfo{journal}{Nucl. \ Instr. \ Meth. \ Phys. \ Res.
  \ A} \textbf{\bibinfo{volume}{893}}, \bibinfo{pages}{138}
  (\bibinfo{year}{2018}).

\bibitem[{\citenamefont{Samanta et~al.}(2018)\citenamefont{Samanta, Das,
  Bhattacharjee, Chatterjee, Raut, Ghugre, Sinha, Garg, Neelam, Kumar
  et~al.}}]{Sam18}
\bibinfo{author}{\bibfnamefont{S.}~\bibnamefont{Samanta}},
  \bibinfo{author}{\bibfnamefont{S.}~\bibnamefont{Das}},
  \bibinfo{author}{\bibfnamefont{R.}~\bibnamefont{Bhattacharjee}},
  \bibinfo{author}{\bibfnamefont{S.}~\bibnamefont{Chatterjee}},
  \bibinfo{author}{\bibfnamefont{R.}~\bibnamefont{Raut}},
  \bibinfo{author}{\bibfnamefont{S.~S.} \bibnamefont{Ghugre}},
  \bibinfo{author}{\bibfnamefont{A.~K.} \bibnamefont{Sinha}},
  \bibinfo{author}{\bibfnamefont{U.}~\bibnamefont{Garg}},
  \bibinfo{author}{\bibnamefont{Neelam}},
  \bibinfo{author}{\bibfnamefont{N.}~\bibnamefont{Kumar}},
  \bibnamefont{et~al.}, \bibinfo{journal}{Phys. \ Rev.\ C}
  \textbf{\bibinfo{volume}{97}}, \bibinfo{pages}{014319}
  (\bibinfo{year}{2018}).

\bibitem[{rad()}]{rad}
\urlprefix\url{www.radware.phy.ornl.gov}.

\bibitem[{\citenamefont{Palit et~al.}(2000)\citenamefont{Palit, Joshi, Nagaraj,
  Rao, Chintalapudi, and Ghugre}}]{Pal00}
\bibinfo{author}{\bibfnamefont{R.}~\bibnamefont{Palit}},
  \bibinfo{author}{\bibfnamefont{P.~K.} \bibnamefont{Joshi}},
  \bibinfo{author}{\bibfnamefont{S.}~\bibnamefont{Nagaraj}},
  \bibinfo{author}{\bibfnamefont{B.~V.~T.} \bibnamefont{Rao}},
  \bibinfo{author}{\bibfnamefont{S.~N.} \bibnamefont{Chintalapudi}},
  \bibnamefont{and} \bibinfo{author}{\bibfnamefont{S.~S.}
  \bibnamefont{Ghugre}}, \bibinfo{journal}{Pramana}
  \textbf{\bibinfo{volume}{54}}, \bibinfo{pages}{347} (\bibinfo{year}{2000}).

\bibitem[{\citenamefont{Guttormsen et~al.}(1983)\citenamefont{Guttormsen, von
  Grumbkow~andY. K.~Agarwal, Blume, Hardt, Hubel, Recht, Schuler, Kluge, Maier,
  Maj et~al.}}]{Gut83}
\bibinfo{author}{\bibfnamefont{M.}~\bibnamefont{Guttormsen}},
  \bibinfo{author}{\bibfnamefont{A.}~\bibnamefont{von Grumbkow~andY.
  K.~Agarwal}}, \bibinfo{author}{\bibfnamefont{K.~P.} \bibnamefont{Blume}},
  \bibinfo{author}{\bibfnamefont{K.}~\bibnamefont{Hardt}},
  \bibinfo{author}{\bibfnamefont{H.}~\bibnamefont{Hubel}},
  \bibinfo{author}{\bibfnamefont{J.}~\bibnamefont{Recht}},
  \bibinfo{author}{\bibfnamefont{P.}~\bibnamefont{Schuler}},
  \bibinfo{author}{\bibfnamefont{H.}~\bibnamefont{Kluge}},
  \bibinfo{author}{\bibfnamefont{K.~H.} \bibnamefont{Maier}},
  \bibinfo{author}{\bibfnamefont{A.}~\bibnamefont{Maj}}, \bibnamefont{et~al.},
  \bibinfo{journal}{Nucl. \ Phy. \ A} \textbf{\bibinfo{volume}{398}},
  \bibinfo{pages}{119} (\bibinfo{year}{1983}).

\bibitem[{nnd()}]{nndc}
\urlprefix\url{www.nndc.bnl.gov}.

\end{thebibliography}
\end{document}